 \documentclass[final,1p,times]{elsarticle}

\usepackage{lineno}
\usepackage{amssymb, amsfonts}
\usepackage{lipsum}
\usepackage{rotating}
\usepackage{pslatex}
\usepackage{epsfig,soul}
\usepackage[normalem]{ulem}
\usepackage{caption,subcaption}
\usepackage{booktabs,multirow}
\usepackage{graphicx,tikz,graphics,epsfig}
\usepackage{url,CJKutf8}
\usepackage{color,xcolor}
\usepackage{booktabs}
\usepackage{multirow}
\usepackage{setspace} 
\usepackage{physics,amsmath} 
\usepackage{bm}  
\journal{Computer Physics Communications}

\begin{document}  

\begin{frontmatter}  
%%%% Title, authors and addresses 
%%%% use the tnoteref command within \title for footnotes;
%%%% use the tnotetext command for theassociated footnote;
%%%% use the fnref command within \author or \affiliation for footnotes;
%%%% use the fntext command for theassociated footnote;
%%%% use the corref command within \author for corresponding author footnotes;
%%%% use the cortext command for theassociated footnote;
%%%% use the ead command for the email address,
%%%% and the form \ead[url] for the home page: 

 \title{Artificial and Eddy Viscosity in Large Eddy Simulation Part 1: Temporal and Spatial Schemes}
  \author[jsun]{Jing Sun\corref{cor1}}
 \ead{j.sun@rug.nl}  
 \author[jsun]{Roel Verstappen}
 \ead{r.w.c.p.verstappen@rug.nl}  
 \cortext[cor1]{Corresponding author}
 \affiliation[jsun]{organization={University of Groningen, Computational and Numerical Mathematics — Bernoulli Institute}, 
            addressline={Nijenborgh 9, 9747 AG, Groningen}, 
            city={Groningen}, 
            country={The Netherlands}}     
   
\begin{abstract}
%\section*{Abstract}
We propose a novel method to quantify artificial dissipation in large eddy simulation. Here, artificial dissipation is defined as the residual of the discrete turbulent kinetic energy (TKE) equation. This method is applied to turbulent channel flow at $\mathrm{Re_\tau}=180$ and 550 using a minimum-dissipation model within the OpenFOAM framework, employing both symmetry-preserving and standard OpenFOAM discretizations. Our analysis reveals that artificial viscosity can both produce and dissipate TKE. To quantify the interaction between subgrid-scale model and artificial dissipation, we introduce viscous activity parameters $\chi$ and $\chi_{nm}$, which also allow for evaluating the balance between molecular and non-molecular viscosity. Examining temporal discretization schemes, we find that all methods can deliver accurate flow predictions if an appropriate time step is used.  For very small time steps, Forward-Euler is more efficient and accurate than Crank-Nicolson, though a better balance of accuracy and computational efficiency for larger time steps, making it preferable for large-scale applications. On collocated meshes, Van Kan's pressure correction method predicts better results than Chorin's.   Our spatial discretization study shows that the symmetry-preserving scheme outperforms standard OpenFOAM schemes, as excessive artificial errors from spatial discretization can suppress the LES model. For the symmetry-preserving scheme, both eddy and artificial viscosities vary similarly, with eddy viscosity being dominant. The role of artificial viscosity changes with mesh refinement: on coarse meshes, it mainly dissipates TKE, while on finer meshes, it can produce TKE, counteracting eddy viscosity. 
%The most accurate Reynolds shear stress predictions occur when non-molecular dissipation is nearly zero. For accurate mean flow kinetic energy, larger non-molecular dissipation is needed on finer meshes.   
\end{abstract}
\end{frontmatter}

%%\linenumbers 
 
\section{Introduction}   
Large eddy simulation (LES) is a widely utilized technique for studying turbulence, aiding in climate change predictions and various engineering applications, ranging from aerodynamics and cardiology. The success of LES relies on the careful and systematic design of discrete algorithms and sub-grid-scale (SGS) parameterizations rather than assembling them opportunistically \cite{alizadeh2022advances}. Symmetry-preserving discretization preserves the energy of the flow system. In combination of minimum-dissipation models, e.g., it is suitable for systematically investigating turbulence.  At practical resolutions, low-order dissipative discretization methods are often employed in LES resulting into an interference between SGS model and artificial dissipation \cite{CASTIGLIONI2019108843}. This leads to counter-intuitive outcomes such as accuracy degradation with grid refinement or no-model simulations outperforming LES with SGS modeling \cite{vreman1996, ouvrard2010}. Quantifying artificial error is crucial for achieving predictable and 'complete' LES \cite{pope2004ten}.

Several studies have addressed the quantification of the artificial dissipation rate in LES. Prior to 2015, most methods were formulated in spectral space \cite{SCHRANNER201584,castiglioni2015numerical}, limiting their application to periodic computational domains and making them difficult to generalize to complex domains with non-periodic boundary conditions. For such cases, quantifying artificial dissipation rates in physical space is more practical.
Following the work of Domaradzki et al. \cite{domaradzki2003effective, domaradzki2005effective}, Schranner et al. \cite{SCHRANNER201584} introduced a method for quantifying artificial dissipation rates in physical space where complex geometries can occur. This method allows for the quantification of artificial dissipation rates and viscosities at the individual cell level or for arbitrary subdomains of the computational domain, starting from the transport equation for kinetic energy\ in integral form, over a control volume.  
By applying a finite volume spatial discretisation and a generic discretisation in time, a residual arises in the discretized version of the kinetic energy equation.
This residual contains all effects of the truncation error of the numerical scheme, including dissipative and dispersive errors. 
Komen et al. \cite{KOMEN2017565} modified Schranner et al.'s \cite{SCHRANNER201584} approach by quantifying the artificial dissipation using the residual of the transport equation of the turbulent kinetic energy and validated their method with quasi-DNS and no-model-LES simulations in turbulent channel flow.

In this paper, we follow the methodology of Komen et al. \cite{KOMEN2017565} and extend the quantification of the artificial dissipation rate to explicit LES in physical space, focusing on the artificial dissipation at small scales and applicability to complex geometries. Additionally, we analyze the artificial  viscosity in minimum-dissipation models, across various temporal and spatial discretization schemes and resolutions, in fully developed turbulent channel flows. The presented explanation of the mechanisms behind the observed trends is based on an analysis of turbulent kinetic energy budgets. 
 
The novel contributions of this paper include: (1) the quantification method for the artificial dissipation rate and the artificial viscosity, (2) the detailed analysis of the impacts of spatial and temporal schemes on the flow quantities and numerical errors, (3) the impact of eddy dissipation and artificial dissipation on turbulent kinetic energy transport and turbulent characteristics, (4) the viscous activity parameter $\chi$ and $\chi_{np}$ are introduced to accurately quantify the interaction between SGS model contributions and numerical error, as well as to evaluate the balance between physical and non-physical viscosity, and (5) detailed analysis of results based on the TKE budgets to understand the long existing problems, including accuracy degradation with the mesh refinement.

The structure of this paper is as follows: Section 2 defines the quantification method for the artificial dissipation rate and artificial viscosity, minimum dissipation model (QR) of the LES approach, symmetry-preserving discretization, error quantification methods, and the viscous activity parameter. Section 3 presents OpenFOAM numerical simulations on turbulent channel flow, validating computational results against detailed DNS data. Section 4 summarizes the conclusions.

\section{Numerical Schemes}
\subsection{Navier-Stokes Equations}
We consider the simulation of turbulent, incompressible flows of Newtonian fluids. The dimensionless governing equations in primitive variables are written as
\begin{align}
     \partial_t \vb{u} + (\vb{u}\cdot \nabla) \vb{u} - \frac{1}{\mathrm{Re}}\nabla \cdot \nabla \vb{u} + \nabla p  = 0 ,\quad \nabla \cdot \vb{u} = 0
     \label{eq:chap6NS}
\end{align}
where $\vb u$ stands for the velocity, $p$ is the pressure and $\mathrm{Re}$ denotes the Reynolds number.   

\subsection{Symmetry-Preserving Spatial Discretization}
\label{sec:symmetry}
In this section, we briefly describe a conservative discretization of the incompressible Navier-Stokes equations \eqref{eq:chap6NS}. To that end, the symmetry properties of the underlying differential operators are maintained: the convective operator is represented by a skew-symmetric matrix, and the diffusive operator is represented by a symmetric positive-definite matrix \cite{verstappen2003symmetry}. On a general unstructured collocated grid, a conservative discretization method was introduced in Ref. \cite{TRIAS2014246,HOPMAN2025113537,santos2025}.
In line with this approach, and using consistent notation, the method employs five fundamental operators: the diagonal matrices for the cell-centered and staggered control volumes, $\bm{\Omega_c}$ and $\bm{\Omega_s}$, 
the matrix of face normal vectors,  $N_s$, the interpolation operator from cell to face, $\Gamma_{c\rightarrow s}$, and the discrete divergence operator, $\vb M$. 
With these operators defined, the discretization follows directly. The temporal evolution of the spatially discrete collocated velocity vector,  $\mathbf{u_c} \in \mathbb R^n$, is then governed by the following operator-based finite-volume discretization of Eq.\eqref{eq:chap6NS}: 
\begin{equation}
   \bm{\Omega}\frac{\mathrm d\vb{u_c}}{\mathrm dt} +\mathbf{C(u_s)u_c + Du_c  +\bm{\Omega}G_c p_c =0, \quad \quad M_c u_c = 0_c}.
    \label{eq:semiSym}
\end{equation}
Here, $\bm \Omega$ is a diagonal matrix containing the size of control volumes associated with the discrete velocity field, and $\vb{p_c}\in \mathbb R^n$ is the cell-centered pressure scalar field. The dimensions $n$ and $m$ correspond to the number of control volumes and faces in the computational domain, respectively. Subscripts $c$ and $s$ distinguish between cell-centered and face-staggered variables.

The diffusive matrix, $\vb D \in \mathbb R^{m\times m}$, represents the integral of the diffusive flux, $-\nu \nabla\mathbf{ u\cdot n}$, across the faces. 
In analogy to the continuous operator, where $\nabla^2 = \nabla \cdot \nabla$, the discrete diffusive operator is the product of a discrete divergence matrix, $\vb{M_s} \in \mathbb R^{m \times m}$, and a discrete gradient matrix. 
The matrix $\vb M \in \mathbb R^{n\times m}$ represents the face-to-cell discrete divergence operator, while the gradient operator is given by the negative transpose of $\vb M$. The convective matrix, $\vb C (\vb{u_s}) \in \mathbb R^{m \times m}$, is skew-symmetric, ensuring conservation. For further details on the discretization of the diffusive and convective operators, we refer to Ref. \cite{verstappen2003symmetry}.

The discrete Laplacian, $\vb{L_c}$, is defined as $\mathbf{L_c =M_c G_c}$, where $\vb {M_c} \equiv \vb M\Gamma_{c\rightarrow s}$ and $\vb{G_c} \equiv \Gamma_{s\rightarrow c} \vb G$  are the collocated discrete divergence and gradient operators, respectively. 
These operators are constructed from the face-to-cell divergence operator, $\vb M$, and the cell-to-face gradient operator, $\vb G$, combined with the interpolation matrices $\Gamma_{c \rightarrow s}$ and $\Gamma_{s \rightarrow c}$. For details regarding the construction of these discrete operators, see Ref. \cite{TRIAS2014246}.

\subsection{Temporal Discretization}
To maintain the favorable conservation and stability properties in discrete-time, the boundary of the stability domain of the time discretization in Eq. (\ref{eq:semiSym}) should coincide with the imaginary axis, which can only be achieved through an implicit temporal integration \cite{PEROT200058}. Therefore, the Crank-Nicolson scheme is utilized for validating and verifying the artificial dissipation and SGS models. In high Reynolds number flow simulations, however, the implicit method may incur higher computational costs even a large time-step can be used compared to the explicit counterpart. Consequently, explicit methods are often used in practice. Therefore, we include various Runge-Kutta schemes \cite{komen2021symmetry,janneshopmanRKSymFoam} in this study too. 
Yet in an explicit temporal discretization scheme for convection-diffusion equations, the time step  $\Delta t$ is typically constrained by convective stability conditions such as $CFL<1$ (where CFL is the Courant–Friedrichs–Lewy condition or Courant number), and diffusive stability conditions akin to $2\Delta t$ $<$ $\mathrm{Re} \Delta y^2$. In this study, the half-implicit and half-explicit Crank-Nicolson scheme is utilized for validating and verifying the artificial dissipation rate and SGS models, while various Runge-Kutta schemes \cite{komen2021symmetry,janneshopmanRKSymFoam} are employed in the sensitivity analysis of the temporal discretization schemes.

\subsection{Correction of Pressure-Velocity Coupling on Collocated Meshes} 
On a collocated grid the decoupling of the velocity and pressure introduces an additional error (proportional to the third-order derivative of pressure) to the momentum equation. Trias et al. \cite{TRIAS2014246} proposed a solution for this decoupling problem without introducing any non-molecular dissipation. The idea behind this approach is to use a linear shift operator to transform a cell-centered velocity $\vb{u_c}$ into a staggered one $\mathbf{u_s}$ and use an incremental-pressure projection method for the explicit time stepping.  Subsequently, Komen et al. \cite{komen2021symmetry} developed a symmetry-preserving, second-order, time-accurate projection and PISO-based methods to achieve nearly conservative results on collocated grids. The discretized operators in the Navier-Stokes equation, as well as the pressure correction method for both implicit and explicit time stepping methods, are thoroughly detailed in the work by Komen et al. \cite{komen2021symmetry}. 

\subsection{Minimum-Dissipation Model}
The first minimum-dissipation model or QR model is proposed by Verstappen \cite{verstappen2011does}. It is based on the invariants of the rate of strain tensor, and set to switch off with negative eddy dissipation. It is given by 
\begin{equation}
    \label{eq:31}
    \tau(v)= -2\nu_e S(v) = -2 C \delta^2 \frac{max\{r(v), 0\}}{q(v)} S(v),
\end{equation}
where $v$ is the velocity field (after the filtering operation s applied), $\tau(v)$ is the anisotropic part of the sub-grid stress tensor, $\nu_e$ is the eddy viscosity, $S(v) $=$ (\nabla v+ \nabla v^T)/2$ is the symmetric part of the velocity gradient. Finally, $q(v)=\mathrm{tr}(S^2(v))/2$ and $r(v)=-\mathrm{tr}(S^3(v))/3= -\mathrm{det}(S(v))$ are the second and third invariant of strain-rate tensor $S(v)$, respectively.  (see Ref. \cite{verstappen2011does,verstappen2018much,rozema2015minimum} for more details). The characteristic filter width $\delta$ is set equal to the geometric mean of the grid size  
\begin{equation}
    \delta = (\delta x \delta y \delta z)^{1/3},
\label{eq:delta}
\end{equation} 
as suggested by Deardorff et al. \cite{deardorff1970numerical}. Scotti et al. \cite{scotti1993} and Trias et al. \cite{trias2017} provide theoretical support for this choice. The QR model is consistent with the exact subfilter stress tensor on isotropic grids (if no artificial dissipation is introduced), that is the difference between the exact stress consists of a $\mathcal O(\delta^4)$ term. The dissipation due to the QR model is
\begin{equation}
    -\tau(v) : S(v) = 2C\delta^2\frac{max\{r(v),0\}}{ q} S(v):S(v)=2 C\delta^2max\{r(v),0\}.
\end{equation} 

\subsection{OpenFOAM Solver}
The OpenFOAM solver used in the present study is called RKSymFoam \cite{komen2021symmetry,janneshopmanRKSymFoam}.
The RKSymFoam solver's primary algorithm comprises three hierarchical iterative levels in the case of an implicit Runge-Kutta (RK) time discretization, e.g., Backward-Euler and Crank-Nicolson:
(1) The outer loop iterates over each Runge-Kutta stage (see \cite{komen2021symmetry} for details).
(2) The intermediate loop manages to update the nonlinear convective term.  
(3) The inner PISO loop handles the pressure-velocity coupling.
For explicit temporal discretization, e.g.,  Forward-Euler, a single projection step suffices for the pressure-velocity coupling, and there's no need to update the convective term iteratively. 

\subsection{Quantification of the Artificial Dissipation Rate and Artificial Viscosity}  
The transport equation for the subgrid turbulent kinetic energy $k=\frac{1}{2} \overline{u'_iu'_i}$ follows from Navier-Stokes equation \eqref{eq:chap6NS}: 
\begin{align}
\frac{Dk}{Dt}  =&\, \partial_t k \hspace{0.3em} +\bar u_j \partial_j k \nonumber\\
= &-\overline{u'_i u'_j}\partial_j \bar u_i
 -  \overline{u'_i u'_j\partial_j u'_i}  +  \nu\partial_j\partial_j k  - \nu\overline{\partial_j u'_i \partial_ju'_i} \nonumber\\
 &+  \nu_e\partial_j\partial_j k - \nu_e\overline{\partial_j u'_i \partial_ju'_i} - \overline{u'_i\partial_ip'},
\label{eq:ksgs} 
\end{align} 
likewise, the time-rate of change of the discrete turbulent kinetic energy $\tilde k = \frac{1}{2}\overline{\vb u_c'^T \Omega \vb u_c' }$ can be derived from Eq.\eqref{eq:semiSym}. This yields 
\begin{align}
    \frac{\tilde D\tilde k}{\tilde Dt}  =&\, \partial_t \tilde k \hspace{0.3em} +\vb {C(\bar u_s)} \tilde k \nonumber\\
    =&\underbrace{-\hspace {0.1em}\overline{\vb u_c' \vb u_c'^T } \vb G_c \vb {\bar u_c}}_{\mathcal P_k}
     \hspace {0.15em}\underbrace{- \hspace {0.1em}\overline{\vb u_c'  \vb u_c'^T \vb G_c \vb u_c'}}_{\mathcal T_k} 
     \hspace {0.15em}\underbrace{+\hspace {0.1em}\vb D \tilde k}_{\mathcal D^\nu_k}
     \hspace {0.15em}\underbrace{- \hspace {0.1em}\nu \vb G_c \vb u_c'\vb G_c \vb u_c'}_{\epsilon_k^\nu} \nonumber\\
     &\hspace {0.15em}\underbrace{+\hspace {0.1em}\vb D_e \tilde k}_{\mathcal D_k^{sgs}} 
     \hspace {0.15em}\underbrace{-\hspace {0.1em}\nu_e \vb G_c \vb u_c' \vb G_c \vb u_c'}_{\epsilon^{sgs}_k} 
     \hspace {0.15em}\underbrace{-\hspace {0.1em} \overline{\vb u_c' \vb G_c \vb p_c'}}_{\mathcal D_k^p},
\label{eq:kdiscrete}
\end{align}
where $\tilde D/\tilde Dt \approx D/Dt$ is the discrete material derivative. The right-hand-side of Eq.\eqref{eq:kdiscrete} is splited into several terms. From left to right: production $\mathcal {P}_k$, turbulent transport $\mathcal {T}_k$, molecular diffusion $\mathcal {D}_k^v$, molecular dissipation $\epsilon^{\nu}_k$,  eddy diffusion $\mathcal{D}^{sgs}_k$, eddy dissipation $\epsilon^{sgs}_k$ and pressure diffusion $\mathcal {D}_k^p$, respectively. These terms are approximated numerically. To avoid the explicit use of fluctuations and to avoid taking derivatives in the post-processing, the terms in the right-hand-side of the Eq.\eqref{eq:kdiscrete} are rewritten as follows. 

The production of the Reynolds stress term becomes 
\begin{equation}
    \mathcal {P}_k = -\overline{\vb u_c' \vb u_c'^T } \vb G_c \vb {\bar u_c}
	= - (\overline{\vb u_c \vb u_c^T}-\vb {\bar u_c} \vb {\bar u_c^T})  \overline{\vb G_c  \vb u_c}
	\label{eq:prod}
\end{equation}
This term is generally positive, and hence it is a source in the $\tilde k$ transport equation.  The action of the mean velocity gradients working against the Reynolds stresses removes kinetic energy from the mean flow and transfers it to the fluctuating velocity field \cite{Pope_2000}. Based on the observations, only the symmetric part of the velocity gradient tensor and the anisotropic part of the Reynolds stress tensor affect the production $\mathcal{P}_k$. With the turbulent-viscosity hypothesis, the production takes the form $\mathcal{P}_k = 2\nu_e \bar S_{ij}\bar S_{ij}$.

The molecular and eddy dissipation of turbulent kinetic energy are defined by  
\begin{align}
   &\epsilon^{\nu}_k= -\nu \vb G_c \vb u_c'\vb G_c \vb u_c'
=- \nu \overline{\vb G_c \vb u_c \vb G_c \vb u_c} + \nu\overline{\vb G_c \vb u_c}\hspace {0.2em} \overline{\vb G_c \vb u_c},\\
    &\epsilon^{sgs}_k =- \nu_e \vb G_c \vb u_c' \vb G_c \vb u_c'
=- \overline{\nu_e\vb G_c \vb u_c \vb G_c \vb u_c} + \overline{\vb G_c \vb u_c}\hspace {0.2em} \overline{\nu_e\vb G_c \vb u_c} ,
\end{align}
respectively.  They are sinks of turbulent kinetic energy. The fluctuating velocity gradient $\vb G_c \vb u_c'$ works against the fluctuating deviatoric stresses $(\nu+\nu_e)s$ transforming kinetic energy into internal energy, where the fluctuating rate of strain $s$ is approximated by $\frac{1}{2}\left(\vb G_c \vb u_c' + (\vb G_c \vb u_c')^T\right)$. (Note that the resulting rise in temperature is almost always negligibly small.)  

The turbulent transport is expressed as 
\begin{align}
    \mathcal{T}_k  = &- \overline{\vb u_c' \vb u_c'^T \vb G_c \vb u_c'} \nonumber \\
	=&-\overline{\vb u_c\vb u_c^T\vb G_c \vb u_c} + \vb{\bar u_c} \overline{\vb u_c^T \vb G_c \vb u_c} + \vb {\bar u_c^T} \overline{\vb u_c \vb G_c \vb u_c} + \overline{\vb u_c\vb u_c^T}\overline{\vb G_c \vb u_c} - 2\vb{\bar u_c} \vb{\bar u_c^T} \overline{\vb G_c \vb u_c} \nonumber\\
	=& \,\mathcal C_k - \mathcal P_k.
\end{align}
The rate of energy transfer from the large scales (energy-containing subrange) determines the constant rate of energy transfer through the inertial subrange; hence the rate at which energy leaves the inertial sub-range and enters the dissipation range is equal to the dissipation rate $\epsilon^{\nu}_k$ \cite{Pope_2000}.

The molecular diffusion $\mathcal D_k^\nu $ and eddy diffusion $\mathcal D_k^{sgs}$ are given by  
\begin{align}
&\mathcal D_k^\nu  = \vb D \tilde k
= \frac{1}{2}\overline{\vb D \vb u_c \vb u_c} - \overline{\vb u_c} \overline{\vb D \vb u_c} - \nu\overline{\vb G_c \vb u_c}\hspace{0.2em}\overline{\vb G_c \vb u_c} = \mathcal{L}^{\nu}_k - \epsilon^\nu_k,\\
&\mathcal D_k^{sgs}  =\vb D_e \tilde k
= \frac{1}{2}\overline{\vb D_e \vb u_c \vb u_c} - \overline{\vb u_c}\hspace{0.2em} \overline{\vb D_e \vb u_c} - \overline{\vb G_c \vb u_c}\hspace{0.2em}\overline{\nu_e\vb G_c \vb u_c} = \mathcal{L}^{sgs}_k - \epsilon^{sgs}_k,
\end{align}
respectively. Therefore, it seems easiest to express molecular diffusion as $\mathcal{L}^{\nu}_k - \epsilon^{\nu}_k$ and eddy diffusion as $\mathcal{L}^{sgs}_k - \epsilon^{sgs}_k$, where both $\mathcal{L}^{\nu}_k$ and $\mathcal{L}^{sgs}_k$ can be approximated using cell-centered discretizations of the velocity gradient. 
  
The rate of production of turbulent kinetic energy by the subgrid model is 
\begin{equation}
\mathcal{P}_{sgs} = - \tau_{ij}:\bar S_{ij} = 2 \overline{\nu_e S_{ij}}:\bar S_{ij}\geq 0.
\end{equation}
This term $\mathcal{P}_{sgs}$ acts as a sink ($-\mathcal{P}_{sgs}$) in the transport equation of the mean kinetic energy and as a source ($+\mathcal{P}_{sgs}$) in the transport equation for the turbulent kinetic energy. It represents, therefore, the rate of energy  transfer from the filtered motions to the residual motions.

The turbulent transport $\mathcal T_k$, the pressure diffusion $\mathcal D_k^p$, the viscous diffusion $\mathcal D_k^v$ and the eddy diffusion $\mathcal D_k^{sgs}$  are transport, or redistribution, terms. These terms do not introduce a net increase or decrease of the mean turbulent kinetic energy, hence no increase or decrease of the subgrid kinetic energy. Therefore, it can be concluded that: 1) the total integrated turbulent transport, pressure diffusion, viscous diffusion and subgrid diffusion rate of $\tilde k$ for the channel flow domain are equal to zero, and 2) the total integrated production rate of $\tilde k$ equals the total integrated dissipation rate of $k$ for the channel flow domain \cite{nieuwstadt2016introduction}. 

Furthermore, for the fully developed statistically stationary turbulent channel flow in the present study, the terms in energy balance equation Eq.\eqref{eq:kdiscrete} are a function of the wall normal coordinate only, and the time rate change $\partial_t \tilde k$ and convective transport $\bar u_j \partial_j \tilde k$ of $\tilde k$ are zero. As a result, the distribution of artificial dissipation rate $\epsilon^{art}_k$ of the turbulent kinetic energy $\tilde k$ across the channel height in statistically steady state conditions is equal to the distribution of the sum of the budget terms of $\tilde k$ \cite{komen2014quasi}
\begin{equation}
-(\mathcal P_k + \mathcal T_k+ \mathcal D_k^\nu    + \mathcal D_k^{sgs} + \mathcal D_k^p + \epsilon^\nu_k + \epsilon^{sgs}_k )= \epsilon^{art}_k.
\label{eq:numDiss}
\end{equation} 
Subsequently, the corresponding artificial viscosity can be computed from the sum of the budget terms of $\tilde k$ divided by the resolved dissipation rate of $\tilde k$ as 
\begin{equation}
\nu_{ art} = \nu \frac{\epsilon^{ art}_k}{\epsilon^{\nu}_k}.
\label{eq:nu_num}
\end{equation}
Note that artificial dissipation (or artificial viscosity) refers specifically to the dissipation of fluctuation energy rather than the total energy. Moreover, this dissipation can be either positive or negative. 

\subsection{Error Quantification}
For a comprehensive and reliable analysis of the computational results, five methods are chosen to quantify the error  in flow quantities.  
The relative difference of $\Psi^+(y)$  between LES and DNS predictions was evaluated as basic error-measure: 
\begin{equation}
\Psi_\varphi(y) = \frac{\varphi^+_{ref}(y) - \varphi^+(y)}{\varphi_{ref}^+(y)},
\label{eq:diff}
\end{equation}
where $\varphi^+(y)$ is chosen to reflect the error in large and small scale motions. For large scale motions, $\varphi^+(y)$ is chosen to be the kinetic energy of the mean flow $\bar E=\frac{1}{2}\bar u \bar u$, and for small scale motions $\varphi^+(y)$ is specified as the shear stress $u'v'$, the turbulent kinetic energy $k$ or the enstrophy $\xi = \frac{1}{2}\overline{|\curl u|^2}$ with vorticity $\omega = \curl u$. The LES data is directly compared with unfiltered DNS data. Statistical averages of $\Psi_\varphi$ (represented as $\overline{\Psi_\varphi}$), integrals of $\Psi_\varphi$ over half of the channel ($\int_0^1 \Psi_\varphi dy$), integrals of the square of $\Psi_\varphi$ over half of the channel ($\int_0^1 \Psi^2_\varphi dy$), maximum $\Psi_\varphi$, and statistical averages of the gradient of  $\Psi_\varphi$ (indicated as $\overline{|\nabla \Psi_\varphi|}$) are utilized. Here, the gradient is computed from the quasi-linear part ($y^+ \le 5$) where the mean velocity is given by $u^+ = y^+ $. 

\subsection{Viscous Activity Parameter}
To quantity the relative importance of the molecular viscosity, the eddy viscosity and the artificial viscosity, we introduce, the following viscous activity parameters:
\begin{equation}
 \chi = \frac{\langle \nu \rangle + \langle \nu_e\rangle+ \langle \nu_{art}\rangle }{\langle \nu \rangle },
\label{eq:eddyNumToMol}
\end{equation}
\begin{equation}
\chi_{nm} = \frac{\langle \nu_e\rangle+ \langle \nu_{art}\rangle }{\langle \nu_e \rangle }.
\label{eq:eddyToNum}
\end{equation}
Here, $\langle \hspace{0.4em}\rangle$ refers to the statistical average. Note that these variables became 
\begin{equation}
   \chi = 
    \begin{cases}
      0 & \quad\text{if}\quad\nu_e + \nu_{art}= -\nu\\
      1 & \quad\text{if}\quad\nu_e + \nu_{art}= 0\\
      2 & \quad\text{if}\quad\nu_e + \nu_{art}= \nu 
    \end{cases}  
    \quad \mathrm{and} \quad
    \chi_{nm} =
    \begin{cases}
      0 &\quad\text{if}\quad \nu_e = -\nu_{art}\\
      1 &\quad\text{if}\quad \nu_{art}=0\\
      2 &\quad\text{if}\quad \nu_e = \nu_{art}
    \end{cases} 
\end{equation}
The ratio $\chi$ contains all the viscosities, whesear the $\chi_{nm}$ is based on the eddy viscosity and the artificial viscosity.
\section{Verification and Validation in Turbulent Channel Flow} 
Turbulent channel flow is one of the most fundamental wall-bounded shear flows and it has been widely used to study the structure of near-wall turbulence \cite{moser1999direct}. The numerical investigations of the QR model combined with symmetry-preserving discretization applied to channel flow are presented, for friction Reynolds numbers $\mathrm{Re}_\tau = 180$ and 550 (based on the half channel width). Implicit and explicit temporal discretization schemes are compared. Sensitivity studies of initial condition, statistical averaging time, temporal resolution, and mesh resolution are conducted. 

\subsection{Physical Parameters}
The Reynolds number, based on the bulk mean velocity and the half-channel width, is expressed as $\mathrm{Re}_b=U_b h/\nu$, where $h$ represents the half-channel width,  $\nu$ is the fluid viscosity, and  $U_b$ denotes the bulk mean streamwise velocity. The results are normalized using wall units, indicated by a plus sign.Thus the friction Reynolds number, wall-normal coordinate, and friction velocity become
\begin{equation}
\mathrm{Re}_{\tau} =\frac{h u_{\tau}}{\nu}, \quad y^+=y\frac{\nu}{u_{\tau}}, \quad \quad \mathrm{and} \quad u^+=\frac{u}{u_{\tau}}.
\end{equation}
The Kolmogorov length-scale, velocity scale and time scale are defined as 
\begin{equation}
\eta =\left(\frac{(\nu + \nu_e)^3}{\epsilon^{\nu}_k}\right)^{1/4}, \quad u_\eta = (\epsilon^{\nu}_k(\nu + \nu_e))^{1/4}, \quad \quad \mathrm{and} \quad t_\eta = \left (\frac{\nu + \nu_e}{\epsilon^{\nu}_k}\right)^{1/2}.
\end{equation}
The integral length-scale is given by
\begin{equation}
L_{int} = \frac{k^{3/2}}{\epsilon^{\nu}_k},
\end{equation}
where $\epsilon^{\nu}_k$ is the dissipation rate of the subgrid kinetic energy $k$. The corresponding non-dimensional parameters are given by 
\begin{equation}
\label{eq:scale1}
{\epsilon^{\nu}_k}^+ =\epsilon^{\nu}_k \frac{\nu}{u_\tau^4}\quad \mathrm{and} \quad k^+ = \frac{k}{u_\tau^2}.  
\end{equation}
The Kolmogorov scales and integral lengthscale are calculated in the quasi-isotropic area ($y^+>90$) where the components of the velocity gradient $\partial u/\partial x$ and  $\partial u/\partial y$ fluctuate to a small extent and are close to zero.

\paragraph{Time-step Size}
The local Courant number is defined as:
\begin{equation}
    CFL = \frac{\Delta t \sum_{f}|u_f|}{2\Omega},
\end{equation}
where $\Delta t$ is the time-step size, $|u_f|$ is the velocity magnitude at the cell face, $\Omega$ is the cell volume, and $\sum_f$ denotes the summation over the cell faces.

\paragraph{Computation of Statistic of Quantities}
The flow is initialized, then develops in time until a statistic steady state is reached, after which the statistics are accumulated. The flow through time (FTT) is defined as:
\begin{equation}
    FTT = \frac{L_x}{U_b},
\end{equation}
where $L_x$ is the streamwise length of the channel and $U_b$ is the bulk velocity employed as a momentum source.

\subsection{Computational Domain} 
\begin{figure}[h!]
    \centering
    \includegraphics[trim= 160 310 260 10, clip, width = 0.48\linewidth]{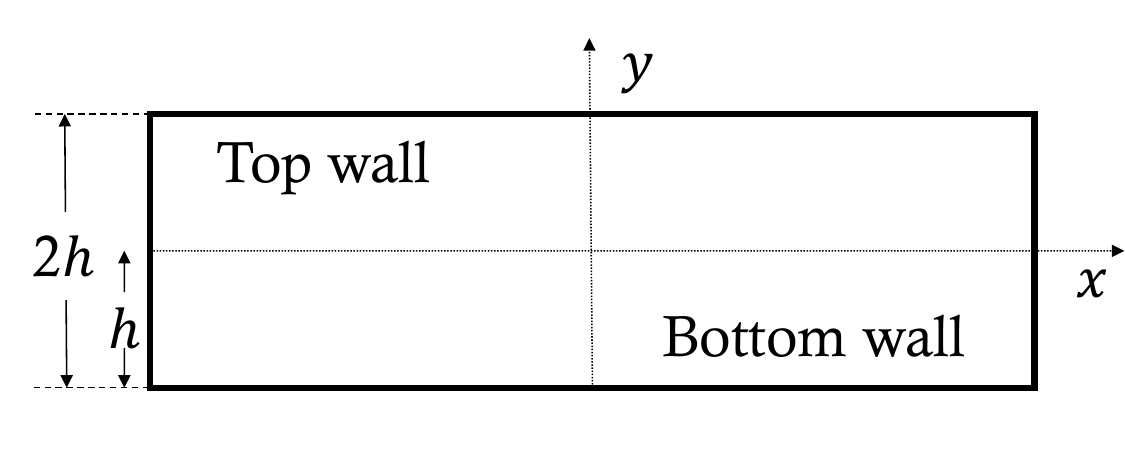}
    \includegraphics[trim= 160 20 260 230, clip, width = 0.48\linewidth]{channelgeometry.pdf} 
    \caption{Numerical domain in xy-plane(left) and xz-plane (right)}
    \label{fig:channel1}
\end{figure} 
The Cartesian coordinate system is shown in Figure \ref{fig:channel1}. The x-axis coincides with the direction of the mean flow and is referred to as the streamwise direction. The y-axis is the wall-normal direction. The z-axis is orthogonal to both x- and y-axis and is called the spanwise direction. The variation of the mean velocity and the Reynolds stress components in the wall-normal direction are matters of interest to researchers and engineers. Many DNS calculations have been carried out and, have produced a considerable amount of informative data. 

In various reference DNS analyses, different computational domain sizes have been employed. (See \cite{vreman2014comparison}, \cite{moser1999direct}, \cite{kim1989transport}, \cite{tiselj2012dns}, \cite{juan2003spectra}, and \cite{kawamura1998dns}, e.g.). An evaluation presented in \cite{komen2014quasi}, Figure 5, suggests that discrepancies among the considered reference DNS databases are generally within approximately $1.5\%$ and $4\%$ for the mean and RMS velocity profiles, respectively, at $\mathrm{Re}_\tau=180$. 
 Given these negligible differences, we have opted to utilize the computational domain of $2\pi\times 2\times\frac{4}{3}\pi$ from Vreman and Kuerten \cite{vreman2014comparison} for channel flow at $\mathrm{Re}_\tau =180$ in the present study.  For simulations at $\mathrm{Re}_\tau=550$, a domain of $4\pi\times 2\times\frac{4}{3}\pi$ is applied which is larger than the size of $2\pi\times 2\times \pi$ in Ref \cite{moser1999direct}.

Fully developed channel flow is homogeneous in the streamwise and spanwise directions, hence periodic boundary conditions are used in these directions. The boundary conditions on the wall are no-slip for velocity. The mesh distribution is uniform in the streamwise and spanwise directions and stretched in the wall-normal direction (clustered near the walls). The stretching ratio is given by $SR=\Delta y_b/\Delta y_w$, where $\Delta y_b$ and $\Delta y_w$ represent the lengths of the cell at the channel center and adjacent to the wall, respectively.
The velocity field is initialized using the results from no model LES after 100FTTs. In this way, much fewer time steps are needed before starting the averaging process. The bulk velocity and molecular viscosity are pre-set such that the $\mathrm{Re}_\tau\approx180$ or $\mathrm{Re}_\tau\approx550$. The friction velocity $u_\tau$ is calculated as $u_{\tau}=\sqrt{\tau_{w}/\rho}$, where $\tau_w = \nu \partial{u}/\partial {y}$ is the wall shear stress.  
 
\subsection{Results and Discussion} 
The results at Reynolds numbers $\mathrm{Re}_\tau$$\approx 180$ and 550 are compared to the DNS data from Vreman and Kuerten \cite{vreman2014comparison} and Moser and Lee \cite{lee2015direct}, respectively.
We have conducted a sensitivity analysis on the CFL number, initial conditions, and statistical averaging time, effectively minimizing the artificial dissipation to a level below $\epsilon^{art}_k<0.15\%$.

\subsubsection{Impact of Temporal Discretization Schemes}
\begin{figure}[bh!]
	\centering
	\includegraphics[width=0.32\linewidth]{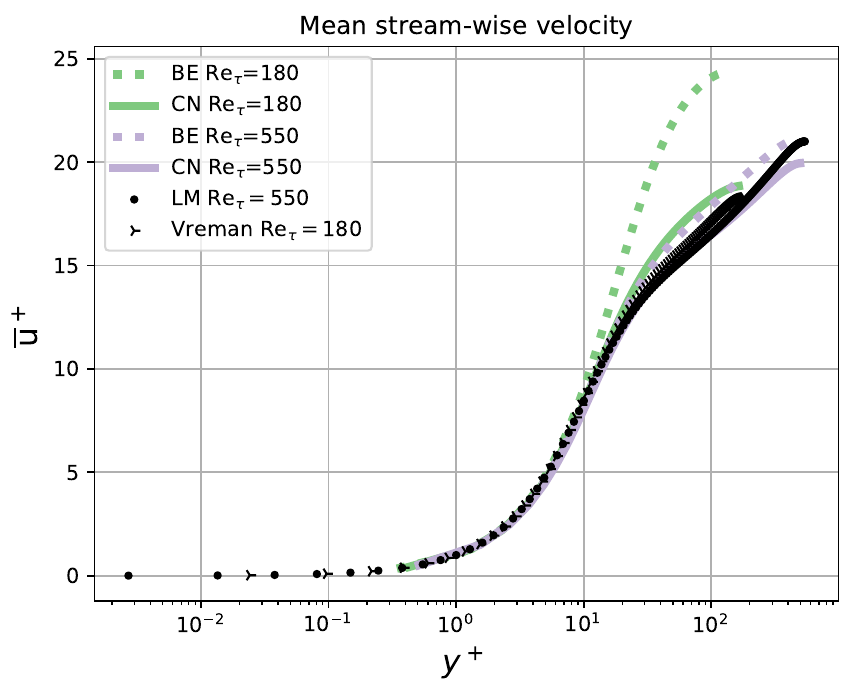}
	\includegraphics[width=0.32\linewidth]{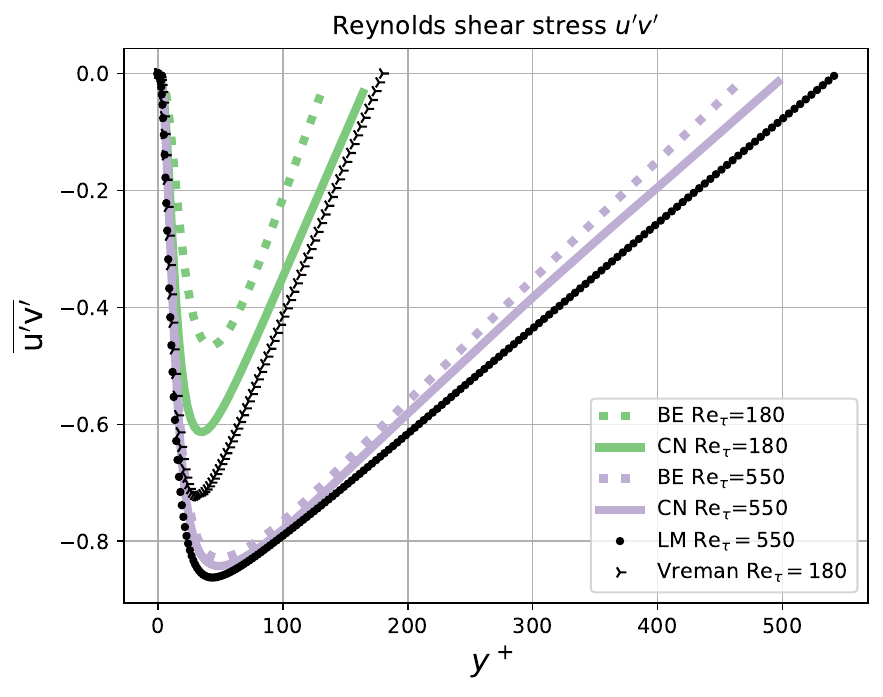}
	\includegraphics[width=0.32\linewidth]{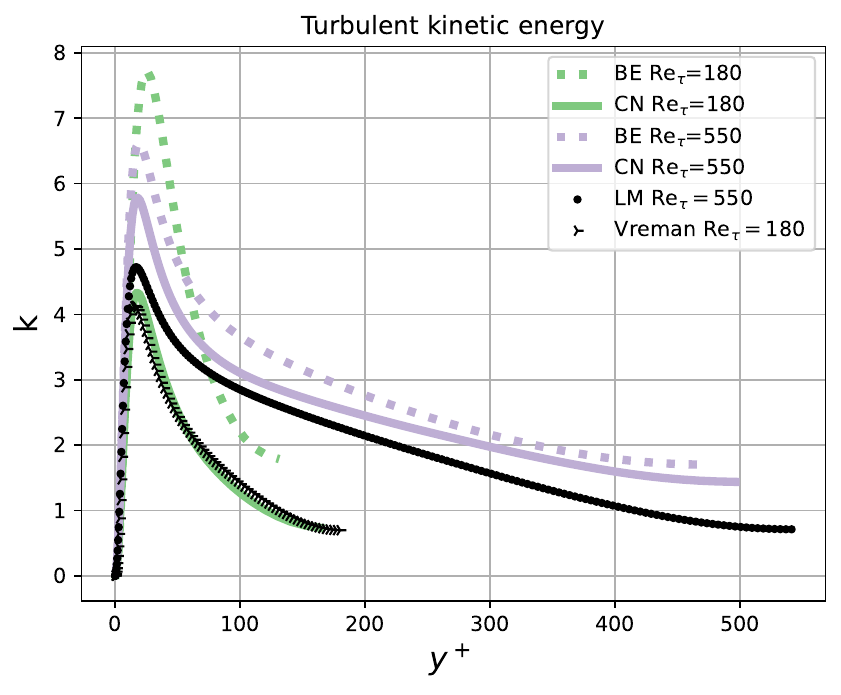}
	\captionsetup{font={footnotesize}}
	\caption{Comparison of the Backward-Euler (BE) and Crank-Nicolson (CN) schemes for channel flow simulations at $\mathrm{Re} = 180$ and 550, with black symbols representing DNS reference data.}
	\label{fig:ddtRe}
\end{figure}

Simulations at two different Reynolds numbers, $\mathrm{Re}_\tau = 180$ and 550, using various time integration schemes have been conducted (see Table \ref{tab:ddt1} for an overview).  The results, shown in Figure \ref{fig:ddtRe}, demonstrate that Backward-Euler provides less accurate predictions of basic flow quantities than Crank-Nicolson at both $\mathrm{Re} = 180$ and 550. 
Here, a symmetry-preserving spatial discretization is applied on a mesh with $168\times96\times96$ grid points and the QR model with coefficient of $C=0.101$ is applied.
\begin{table}[bh!] 
\footnotesize
\centering
\begin{tabular}{ccccccccc} 
\toprule
\multirow{2}{*}{Scheme} & \multicolumn{3}{c}{Explicit}&\multicolumn{4}{c}{Implicit} \\   
\cmidrule(lr){2-4}  \cmidrule(lr){5-8} 
 & FE & RK3& RK4 &BE&CN  & DIRK2 & DIRK3 \\
\cmidrule{1-4}\cmidrule(lr){5-8}
Temporal order& $\mathcal{O}(\Delta t^1)$  & $\mathcal{O}(\Delta t^3)$ & $\mathcal{O}(\Delta t^4)$ & $\mathcal{O}(\Delta t^1)$ & $\mathcal{O}(\Delta t^2)$ &$\mathcal{O}(\Delta t^2)$ &$\mathcal{O}(\Delta t^3)$  \\ 
\bottomrule
\end{tabular} 
\captionsetup{font={footnotesize}}
\caption{Explicit and implicit schemes used for temporal discretization. FE, BE and CN refer to Forward Euler, Backward Euler and Crank-Nicolson schemes, respectively. DIRK2 and DIRK3 denote second-order and third-order Diagonal Implicit Runge-Kutta schemes, respectively. RK3 and RK4 represent third-order and fourth-order Runge-Kutta schemes, respectively.}
\label{tab:ddt1}
\end{table}  
\begin{figure}[!t]
	\centering
	\includegraphics[width=0.325\linewidth]{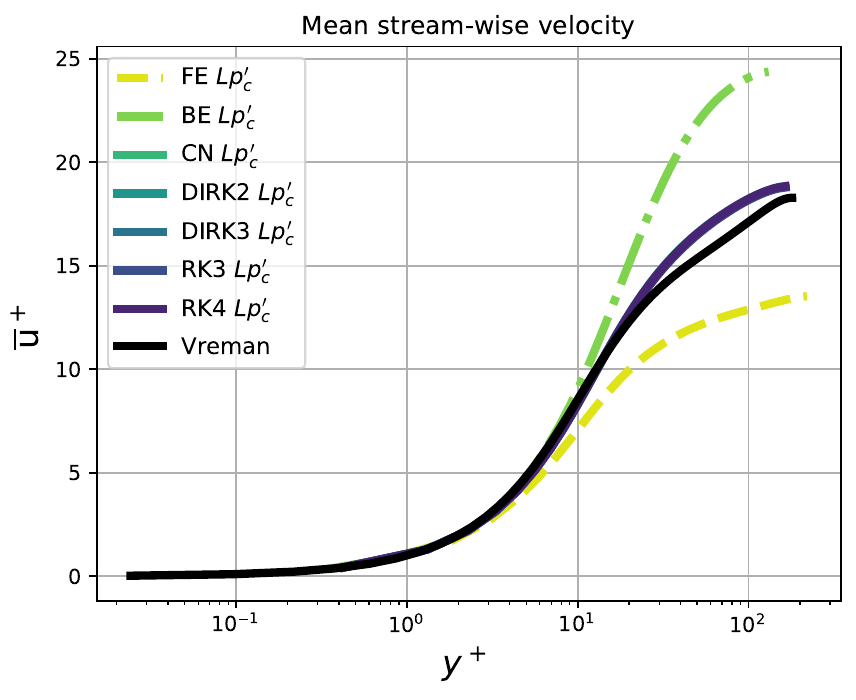} 
	\includegraphics[width=0.325\linewidth]{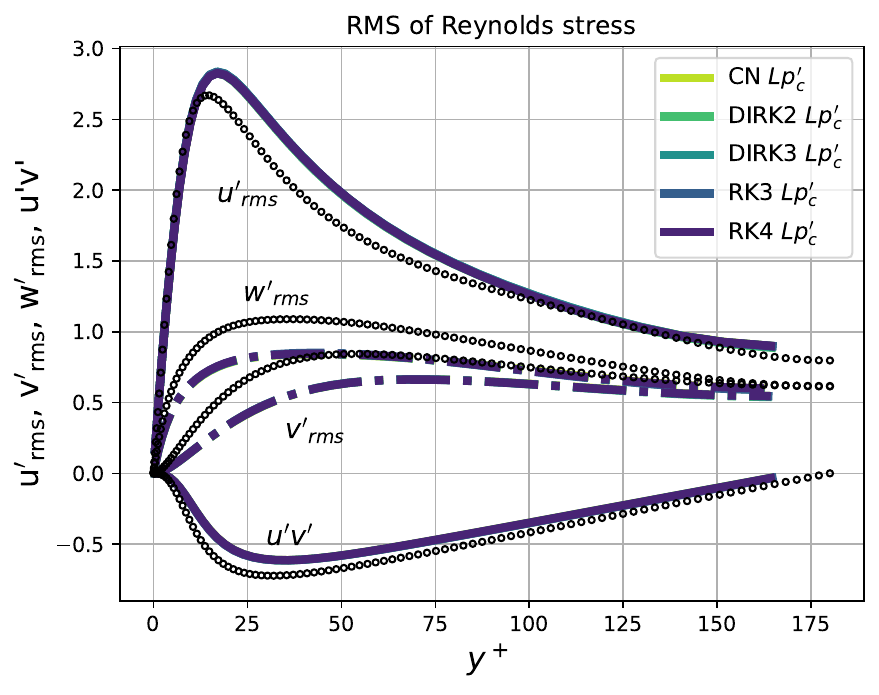}    
	\includegraphics[width=0.325\linewidth]{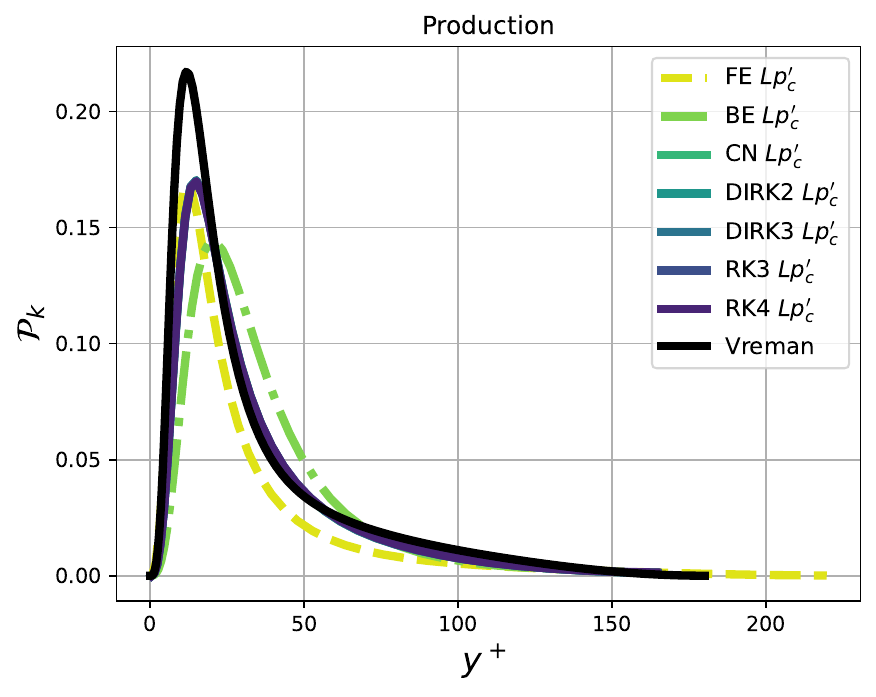}    
	\includegraphics[width=0.325\linewidth]{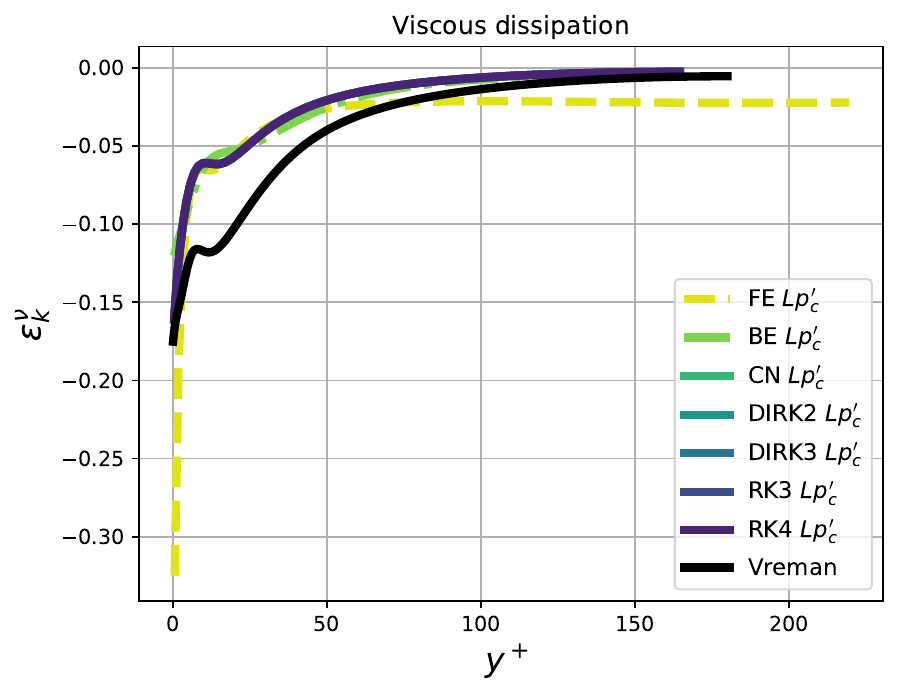}   
	\includegraphics[width=0.325\linewidth]{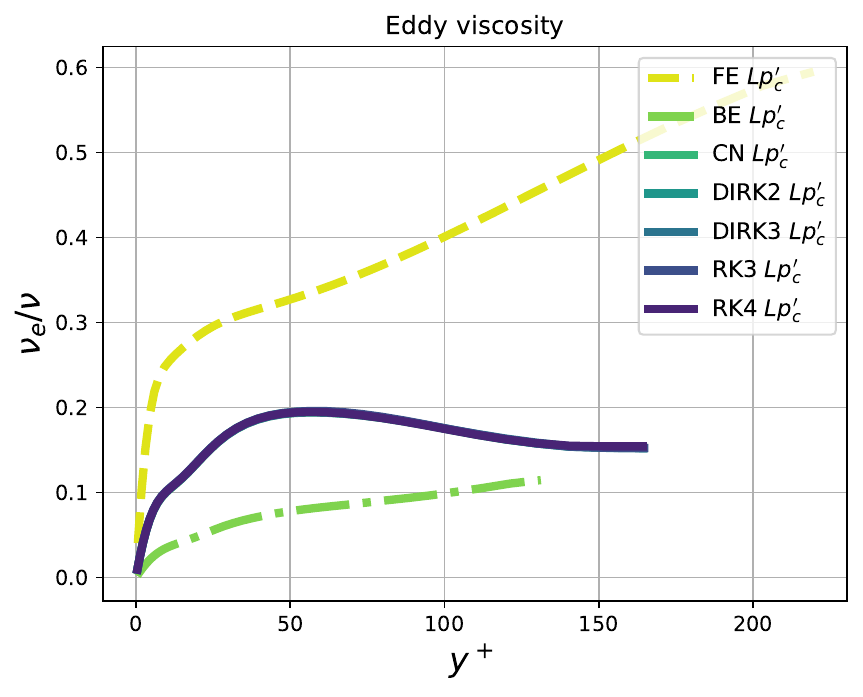}
	\includegraphics[width=0.325\linewidth]{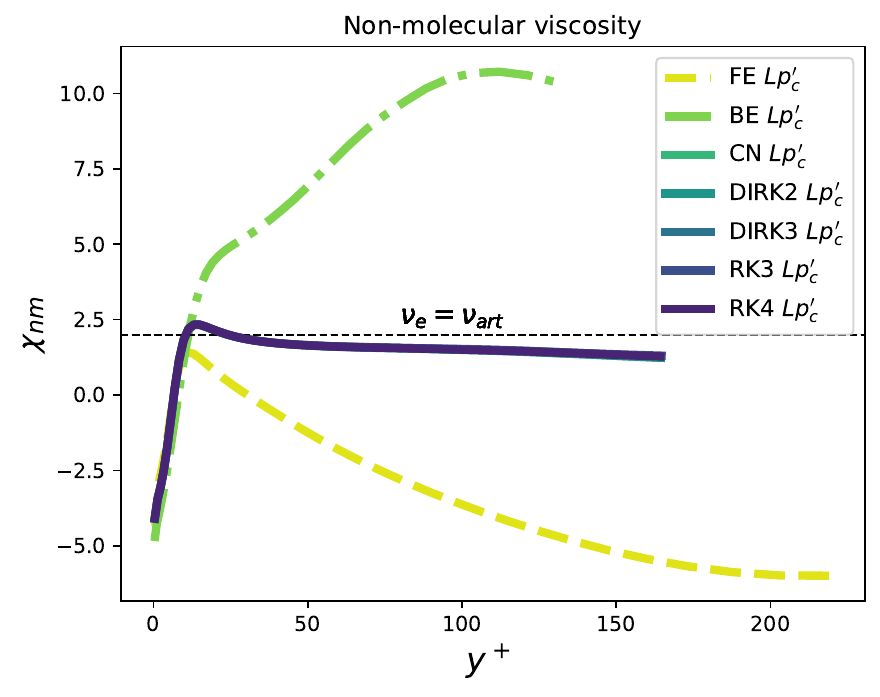}   
	\captionsetup{font = {footnotesize}}
	\caption{Basic flow variables, TKE budget terms, eddy viscosity and viscous parameter versus wall-normal distance in channel flow simulations at $\mathrm{Re}_\tau = 180$, showing the sensitivity of temporal schemes using symmetry-preserving discretization coupled with the QR model ($C = 0.101$) and Van Kan pressure correction.}
	\label{fig:ddt1}
\end{figure}
\begin{table}[h!] 
\footnotesize
\centering
\begin{tabular}{ccccccccccc}
\toprule
Scheme & $\mathrm{Re}_{\tau}$ & $\mathrm{Re}_b$ & $u_{\tau}$ & $u_b$ & $\bar \epsilon^\nu_k$ & $\eta$ & $u_\eta$ & $t_\eta$ & $L_{int}$ \\
\midrule 
FE&228.2&2362.1&0.01088&0.113&0.02227&0.1005&0.1308&0.7684&1330.2\\
BE&136.4&2105.8&0.00650&0.100&0.00759&0.1027&0.0920&1.1163& 434.5\\
CN&170.6&2225.8&0.00813&0.106&0.00397&0.1256&0.0793&1.5837& 234.0\\
DIRK2&170.9&2225.6&0.00815&0.106&0.00391& 0.1260&0.0790&1.5958&230.5\\
DIRK3&170.8&2225.9&0.00815&0.106& 0.00393&0.1259&0.0791&1.5911& 233.3\\
RK3&171.1&2226.1&0.00816&0.106&0.00391&0.1261&0.0790& 1.5965&233.0\\
RK4&170.7&2225.2&0.00814&0.106&0.00396&0.1257&0.0792&1.5865& 231.6\\
\bottomrule
\end{tabular} 
\captionsetup{font={footnotesize}}
\caption{Physical parameters obtained for various temporal discretization schemes. These results are computed using a symmetry-preserving discretization conjugated with the QR model with $C=0.101$ and Van Kan method. CN: Crank-Nicolson; BE: Backward Euler; RK: Runge-Kutta. The results shown are non-dimensionalized as in Eq.\eqref{eq:scale1}. }
\label{tab:ddt2}
\end{table} 

At $\mathrm{Re}_\tau=180$, we consider a range of temporal schemes, encompassing four implicit methods -- Backward-Euler, Crank-Nicolson, DIRK2 (second-order accurate Diagonal Implicit Runge-Kutta), and DIRK3 (third-order accurate Diagonal Implicit Runge-Kutta) -- as well as three explicit discretization techniques, namely Forward-Euler, RK3 (third-order Runge-Kutta schemes), and RK4  (fourth-order Runge-Kutta schemes), see also Table \ref{tab:ddt1}. Both the pressure correction of the first-order Chorin method (referred to as $Lp_c$) and the second-order van Kan method (referred to as $Lp_c'$) are applied to each temporal scheme \cite{komen2021symmetry}. 
A dynamic time step is employed to maintain a maximum CFL number of 0.4, resulting in different mean time steps, $\overline {\Delta t}$, depending on the time integration methods used. These varying time steps directly affect the accuracy of the predicted flow quantities.
The spatial discretization uses a second-order symmetry-preserving scheme \cite{komen2021symmetry}, and the mesh contains $48\times76\times48$ grid points. The subgrid-scale (SGS) model is the QR model \cite{verstappen2011does} with a model coefficient of $C=0.101$. The reference DNS employs a second-order hybrid explicit/implicit three-stage Runge-Kutta method.
The mean streamwise velocity, $\bar{u}$, root mean square of Reynolds stresses $u'u'$, $v'v'$, and $w'w'$, Reynolds shear stress $u'v'$, eddy viscosity $\nu_e$, and the viscous parameter $\chi_{nm}$ are presented in Figure \ref{fig:ddt1}.

From Figure \ref{fig:ddt1}, it is clear that first-order Forward-Euler and Backward-Euler schemes converge, but introduce significant errors. 
For the Forward-Euler method,  $\chi_{nm}<0$ across most of the computational domain, indicating that the artificial viscosity significantly produces turbulent kinetic energy. In other words, the time-integration method fails to dissipate sufficient energy, leading to an accumulation of energy and destabilizing the simulation. In response, the SGS model compensates by introducing substantial dissipation, as reflected in the profile of $\nu_e$. 
Conversely, in the Backward-Euler scheme, $\chi_{nm}>2$ over much of the domain, revealing that the artificial viscosity is considerably larger than the eddy viscosity, resulting in excessive damping of turbulent kinetic energy. This indicates that the time-integration method dissipates too much energy, leading to a suppression of the SGS model, as evident from the reduced eddy viscosity ($\nu_e$) profile. 

For higher-order temporal schemes, including Crank-Nicolson, DIRK2, and DIRK3, and third-order and fourth-order Runge-Kutta schemes, very small differences in the predictions of the flow quantities are found.
These schemes give almost identical predictions in all the flow quantities, including the turbulent kinetic energy and the TKE budget terms. The Reynolds number, Kolmogorov scales, and integral length scale demonstrate close similarity, as indicated in Table \ref{tab:ddt2}.  
The eddy viscosity distribution is consistent across these schemes, increasing sharply from zero near the wall to a peak value of approximately $0.2\nu$ at $y^+ \approx 50$, then gradually decreasing to $0.15\nu$ at the channel center. This distribution is nearly uniform throughout the computational domain.
Additionally, for non-molecular viscosity, $\chi_{nm}$ becomes negative in the near-wall region ($y^+ < 12$), indicating that the magnitude of artificial viscosity exceeds that of eddy viscosity, leading to the production of turbulent kinetic energy. In the region $y^+ > 25$, the viscous parameter falls within the range $1 < \chi_{nm}< 2$, indicating that artificial viscosity and eddy viscosity are comparable, with relatively uniform distributions in this region. 

\paragraph{Analysis of Temporal Resolution and PISO Iterations} 
\begin{figure}[!b]
	\centering 
	\includegraphics[width=0.325\linewidth]{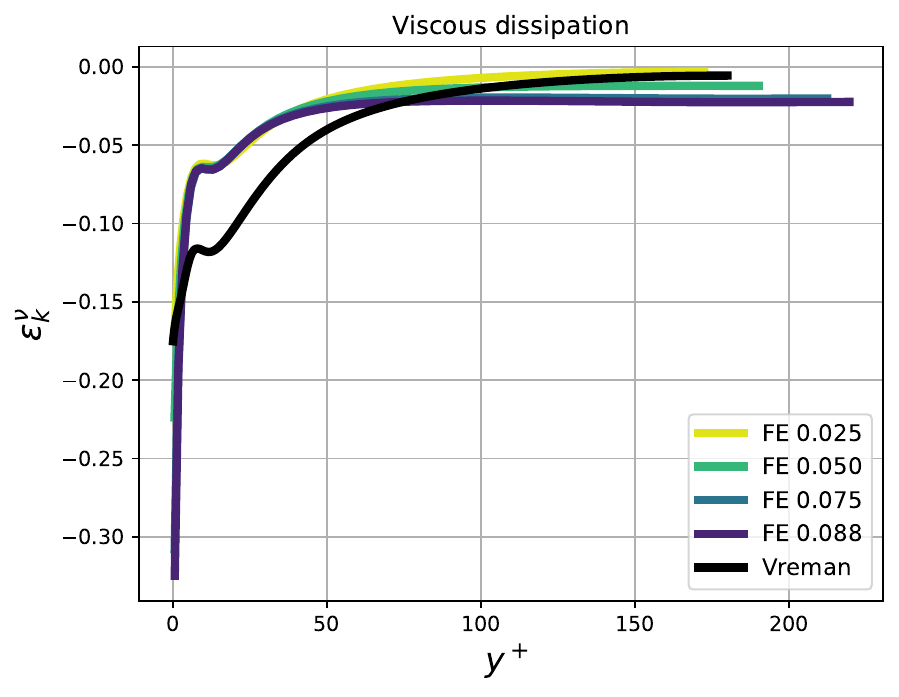}   
	\includegraphics[width=0.325\linewidth]{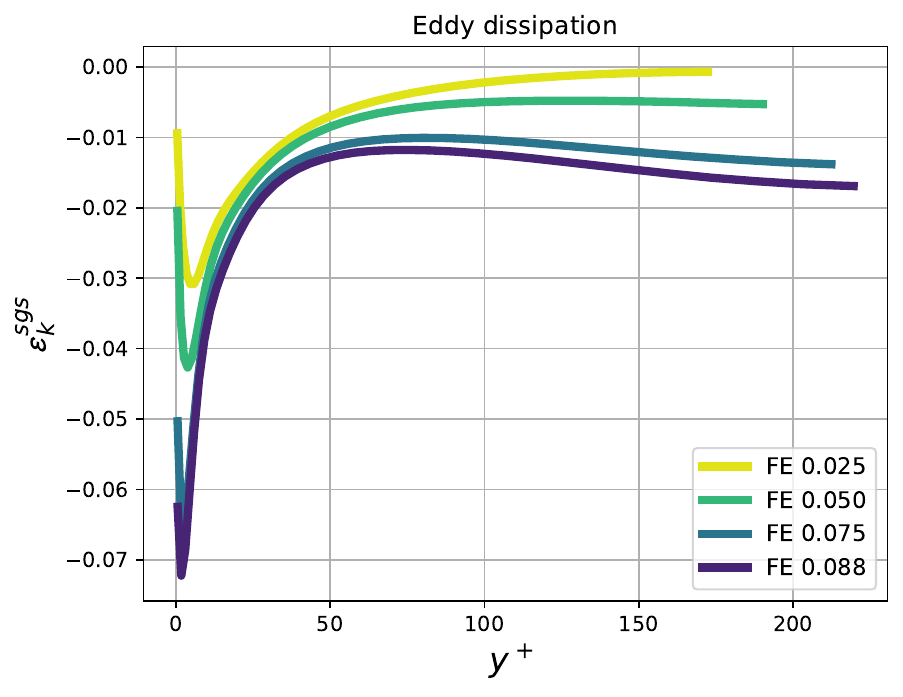}    
	\includegraphics[width=0.325\linewidth]{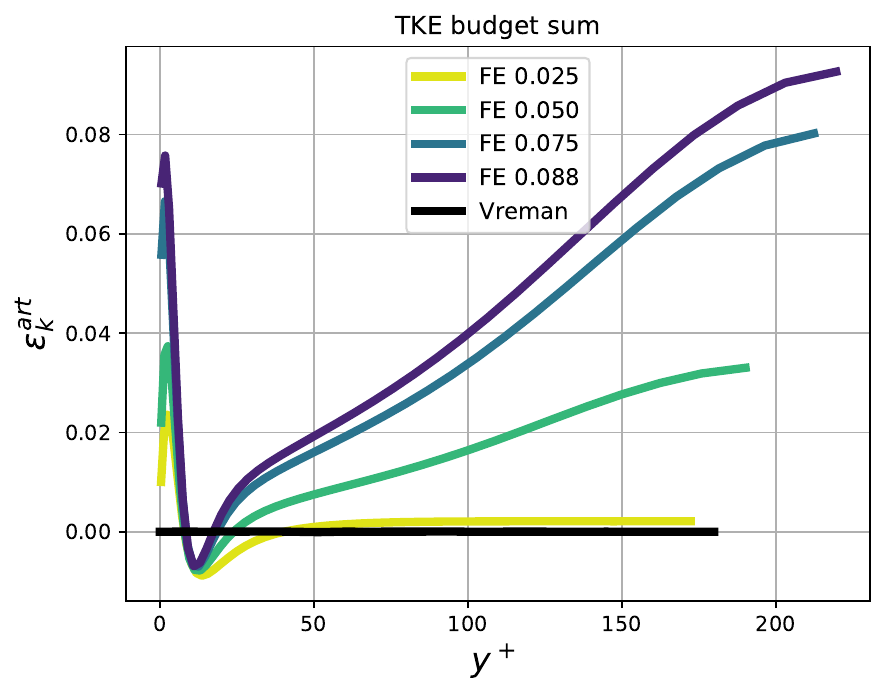}     
	\includegraphics[width=0.325\linewidth]{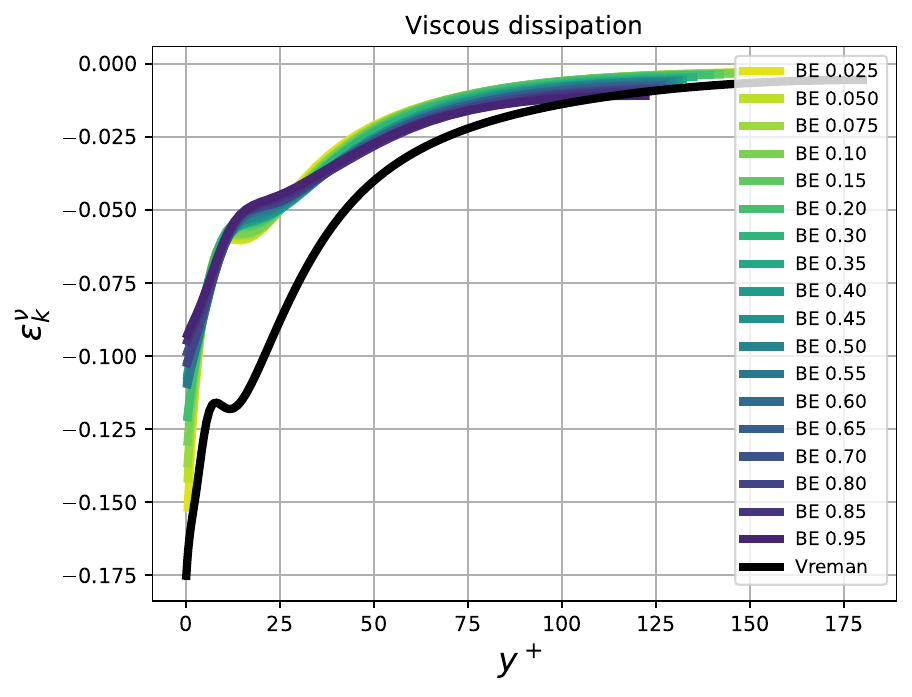}   
	\includegraphics[width=0.325\linewidth]{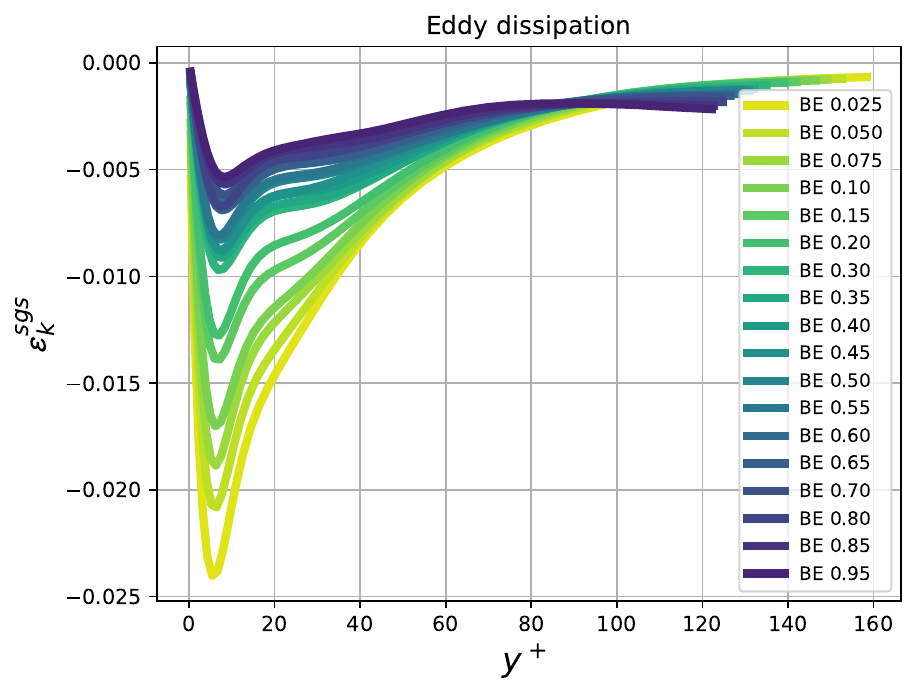}    
	\includegraphics[width=0.325\linewidth]{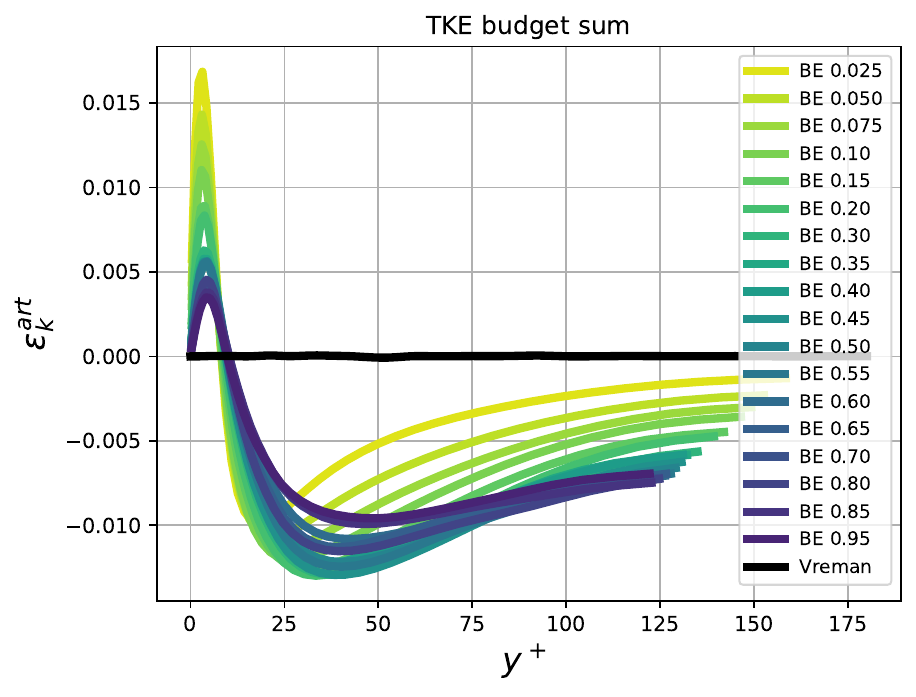}     
	\includegraphics[width=0.325\linewidth]{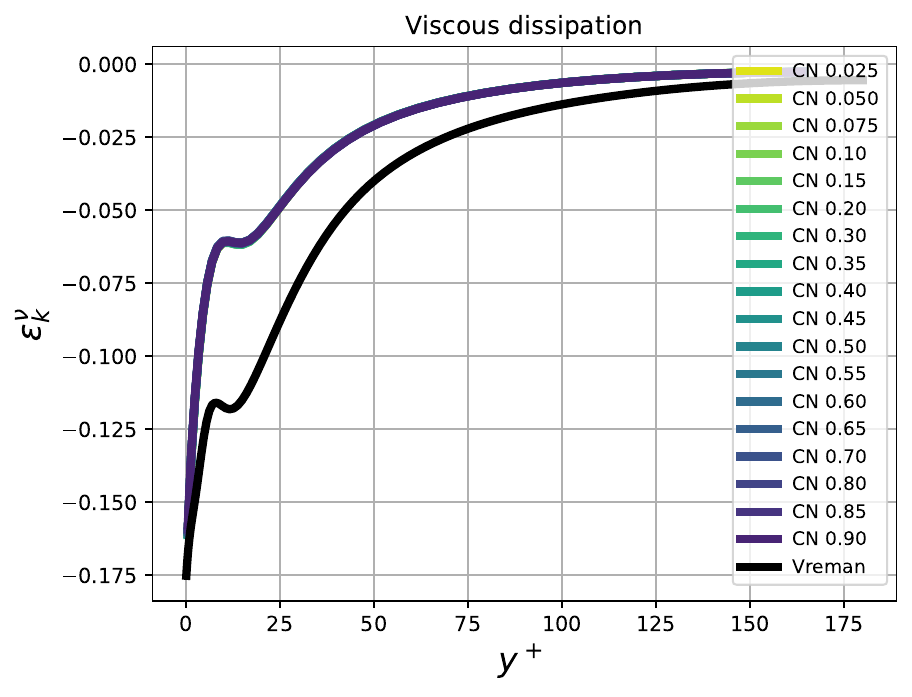}   
	\includegraphics[width=0.325\linewidth]{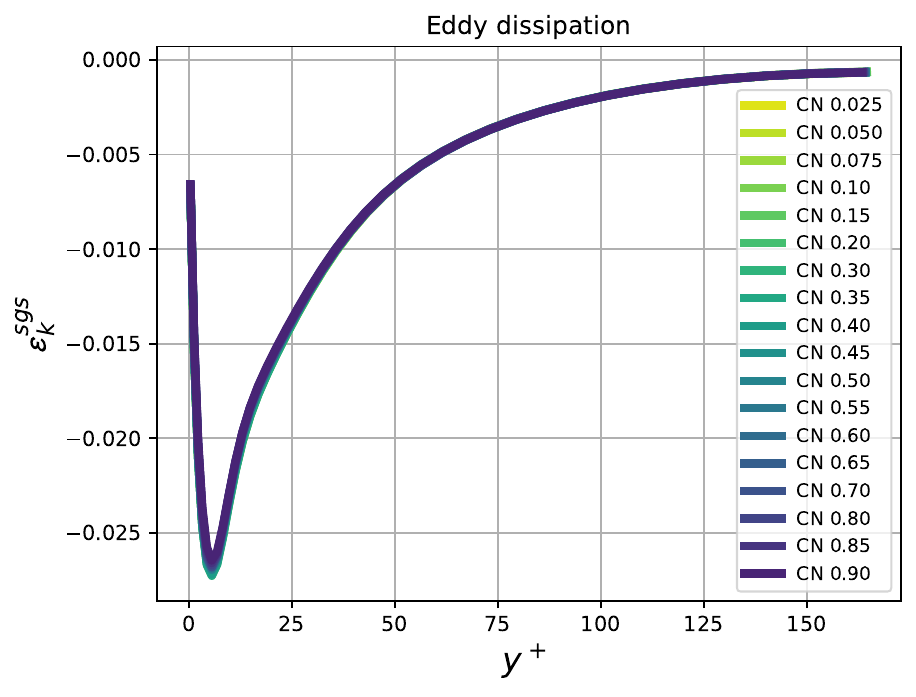}    
	\includegraphics[width=0.325\linewidth]{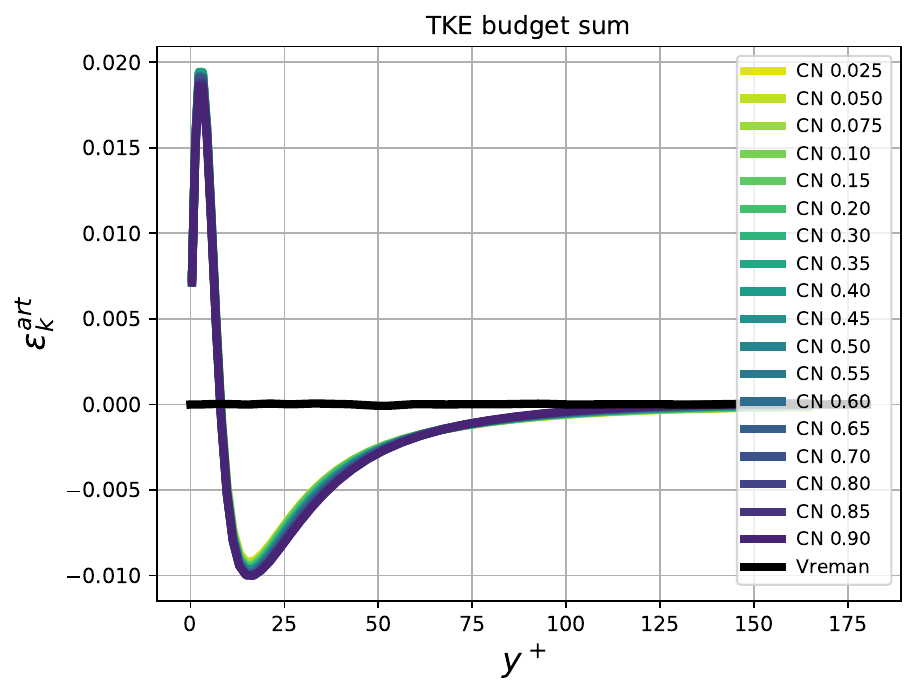}      
	\captionsetup{font = {footnotesize}} 
	\caption{Molecular, eddy and artificial viscosities in channel flow simulations at $\mathrm{Re}_\tau = 180$ with Backward-Euler (BE), Forward-Euler (FE)  and Crank-Nicolson (CN) schemes, for a range of time-steps (using symmetry-preserving discretization and the QR model ($C = 0.101$)).}
	\label{fig:deltaT}
\end{figure} 
\begin{figure}[!t]
	\centering 
	\includegraphics[width=0.43\linewidth]{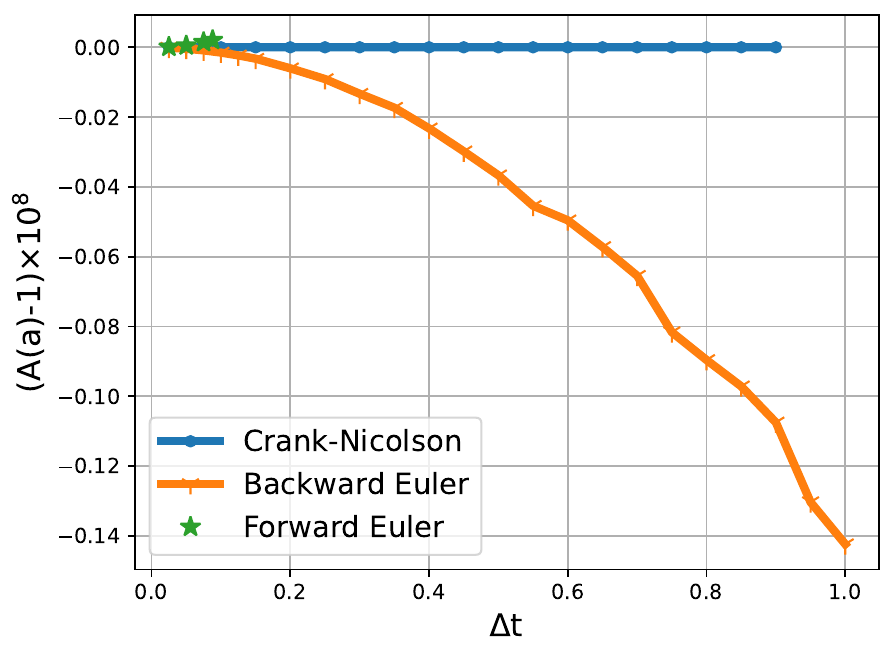}       
	\includegraphics[width=0.45\linewidth]{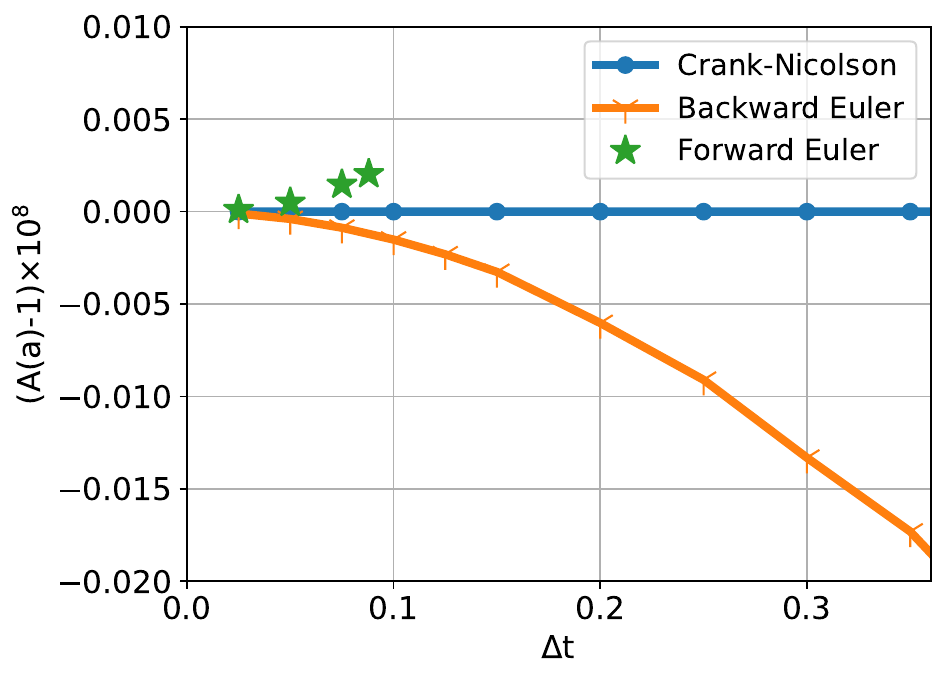}       
	\captionsetup{font = {footnotesize}} 
	\caption{Ratio of the physical to numerical amplification factor as a function of time step for Forward-Euler, Backward-Euler, and Crank-Nicolson schemes in a channel flow simulation at $\mathrm{Re}=180$. Right: a zoomed-in view of the small time-step region.}
	\label{fig:amplification}
\end{figure} 

A  analysis of the time step and PISO iterations was conducted as part of a broader investigation into temporal discretization schemes. The Forward-Euler, Backward-Euler, and Crank-Nicolson schemes were tested. A fractional-step method is employed to solve the velocity-pressure coupling, using a fixed predictor step of 2 and a corrector step of 4. Two approaches to time step control were employed: first, using a dynamic $\Delta t$ to maintain a maximum CFL number of 0.2, 0.4, 0.6, and 0.8; second, fixing the time step $\Delta t$ in the range of 0.025 to 1.0 while keeping the CFL dynamic. For Backward-Euler, increasing $\Delta t$ beyond 0.95  led to solver divergence.  
For Forward-Euler, simulations converged only when $\Delta t < 0.09$. The PISO iteration study was conducted for a fixed CFL of 0.4, varying predictor steps (1, 2, 3, 4, 8, and 16) and corrector steps (2, 3, 4, 8, and 20).  Figure \ref{fig:deltaT} presents the results of varying time step for Forward-Euler (FE), Backward-Euler (BE), and Crank-Nicolson (CN) time integration schemes.

Next we consider the amplification factor of the time-integration schemes, where the eigenvalue is given by $\lambda = -\nu - \nu_e$, where $\nu_e$ is the maximum value of the instantaneous eddy viscosity. The amplification factor $a$ for each time integration is given by: 
\begin{align}
&a_{FE} = 1+ \lambda \Delta t  \quad \quad \hspace{0.2em} \mathrm{Forward-Euler} \\
&a_{BE} = \frac{1}{1-\lambda \Delta t} \quad \quad \mathrm{Backward-Euler}\\
&a_{CN} = \frac{1+\frac{1}{2}\lambda \Delta t}{1-\frac{1}{2}\lambda \Delta t}  \quad \hspace{0.4em} \mathrm{Crank-Nicolson}.
\end{align} 
The ratio of the physical to the numerical amplification factor is defined by
\begin{equation}
\left(A(a)\right)^n=\frac{e^{\lambda n\Delta t}}{a^n}. 
\end{equation} 
For the Crank-Nicolson scheme (after 180 FTTs of averaging), the only notable difference appears in the predictor steps, where a single predictor step increases artificial viscosity at $y^+ \approx 25$. The profiles for various CFL numbers and PISO iterations overlap. The amplification factor ratio, shown in Figure \ref{fig:amplification}, approaches unity for Crank-Nicolson, indicating that the time-integration is accurate, even for relatively large time steps.\\
For the Backward-Euler scheme, reducing the time step significantly improves the accuracy of the predictions, and results converge to DNS reference data as $\Delta t$ decreases. Then, the amplification factor ratio approaches 1. As $\Delta t$ increases, the amplification factor decreases, resulting in overly stable simulations and inaccurate predictions of flow quantities due to excessive artificial dissipation. \\
For the smallest time step of $\Delta t = 0.025$, the Forward-Euler scheme yields more accurate flow quantities than the second-order Crank-Nicolson. However, despite this accuracy, Forward-Euler introduces more artificial viscosity than Crank-Nicolson. The artificial viscosity contributes to the production of the turbulent kinetic energy which is the opposite of other methods. For $\Delta t > 0.09$, Forward-Euler leads to unstable simulations. Figure \ref{fig:amplification} demonstrates that increasing the time step causes the amplification factor ratio to exceed 1, indicating insufficient dissipation for stable simulations.

Overall, all time integration schemes can predict flow variables accurately if an appropriate time step is used, but small time steps are computationally expensive. Moreover, computational time per time-step varies between schemes. To optimize time integration, a comparison was made between Backward-Euler and Crank-Nicolson schemes under common conditions. 

For a total simulation time of 200 FTTs (20 FTTs development and statistical averaging over 180 FTTs), corresponding to a physical time of 19010 seconds and a maximum CFL =0.4, the mean time steps were $\overline{\Delta t_{BE}} \approx 0.367$ and $\overline{\Delta t_{CN}} \approx 0.205$. For Crank-Nicolson, $\overline{\Delta t_{CN}} = 0.35$ yields similarly accurate results, allowing for a direct comparison at $\Delta t = 0.35$, too.
After 200 FTTs, the difference of amplification factor ratio between Crank-Nicolson and Backward-Euler at $\Delta t = 0.35$ is of the order of the molecular and eddy viscosity.The Backward-Euler method introduces significantly more dissipation, yielding overly stable simulations with inaccurate flow predictions. The total simulation time on 32 cores for Backward-Euler was 6.3 hours, compared to 4 hours for Crank-Nicolson. For Crank-Nicolson with $\Delta t = 0.9$, the simulation time was further reduced to 1.5 hours, with more accurate results than Backward-Euler.   
In conclusion, the Crank-Nicolson scheme with $0.35<\Delta t<0.9$ is more accurate and less computationally expensive than Backward-Euler using same settings in the current test case. 
 
Similarly, comparing Forward-Euler and Crank-Nicolson we first note that Forward-Euler is unstable if $\Delta t > 0.09$. At the smallest time step, $\Delta t = 0.025$, Forward-Euler achieved the most accurate predictions. The computational time for 180 FTTs with $\Delta t = 0.025$ was 18.6 hours for Forward-Euler and 41.9 hours for Crank-Nicolson. Thus, Forward-Euler is more than twice as fast and more accurate at far small time steps. However, for larger time steps ($\Delta t = 0.9$), Crank-Nicolson achieved similar accuracy in a much shorter time. 
 
In conclusion, at far small time-step, Forward-Euler outperforms  Crank-Nicolson in achieving accurate results and saving computing time.  Crank-Nicolson provides a better balance between accuracy and computational efficiency at larger time steps, making it more suitable for large-scale simulations.  
 
\paragraph{Study of Pressure Corrections for Solving the Checkerboard Problem}
\begin{figure}[!b]
	\centering
	\includegraphics[width=0.315\linewidth]{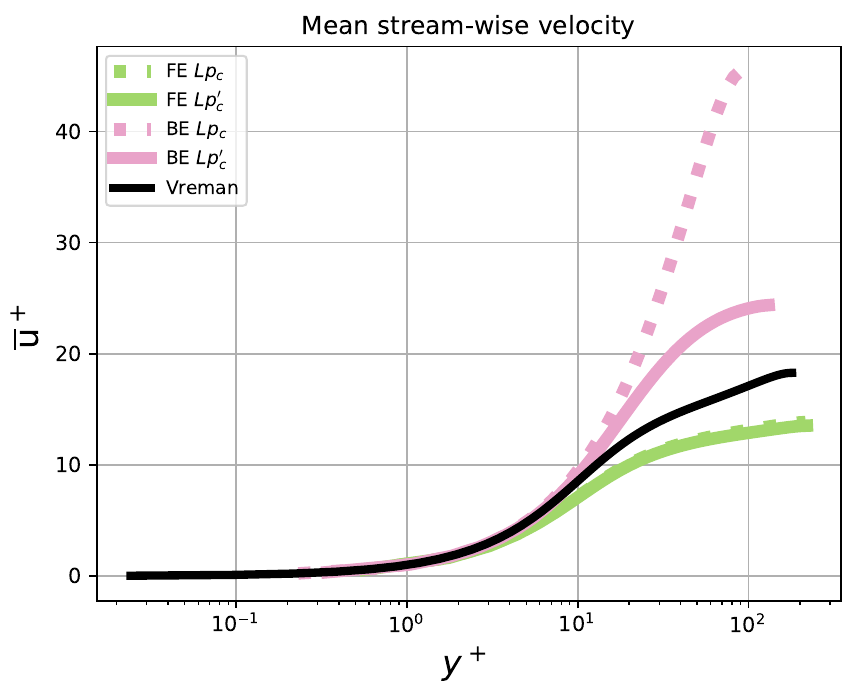}
	\includegraphics[width=0.34\linewidth]{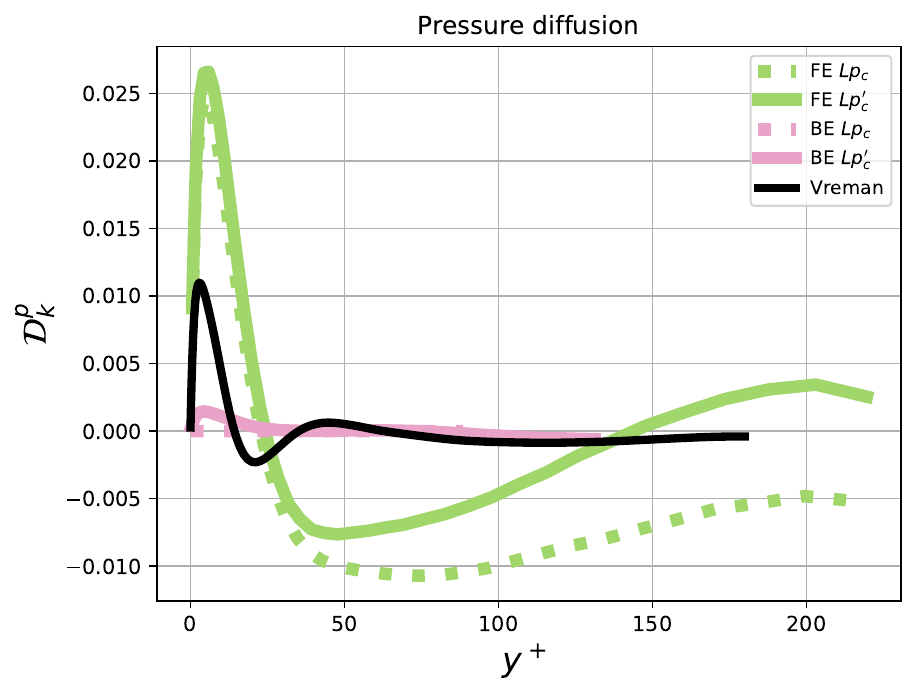} 
	\includegraphics[width=0.325\linewidth]{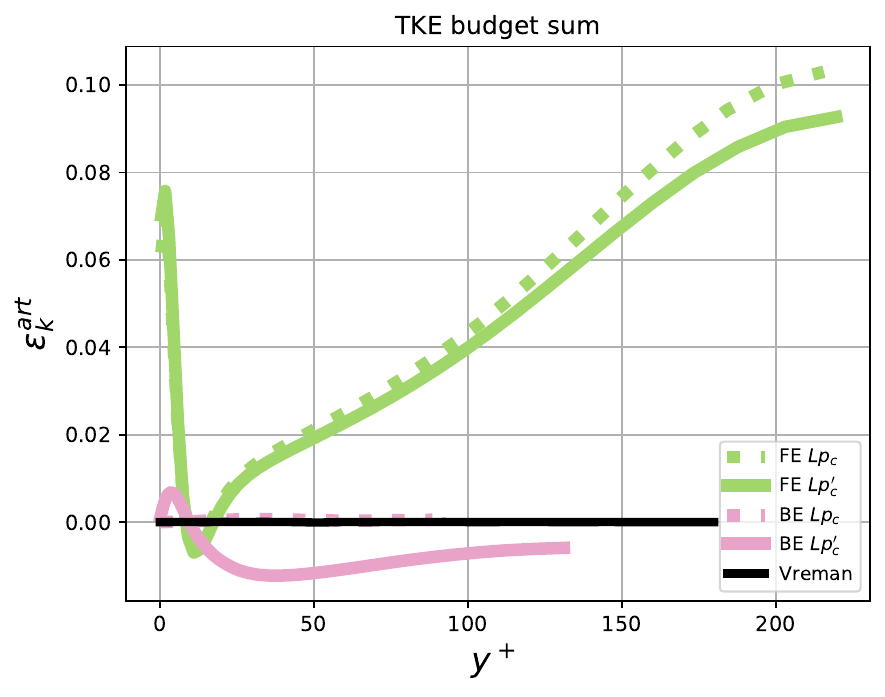}   
	\captionsetup{font = {footnotesize}}
	\caption{Basic flow quantities and change rate of turbulent kinetic energy in channel flow simulations at $\mathrm{Re}_\tau = 180$, using Chorin's and Van Kan's pressure correction methods in combination with the first-order Forward and Backward-Euler schemes.}
	\label{fig:pn_1st}
\end{figure}
\begin{figure}[!b]
	\centering    
	\includegraphics[width=0.32\linewidth]{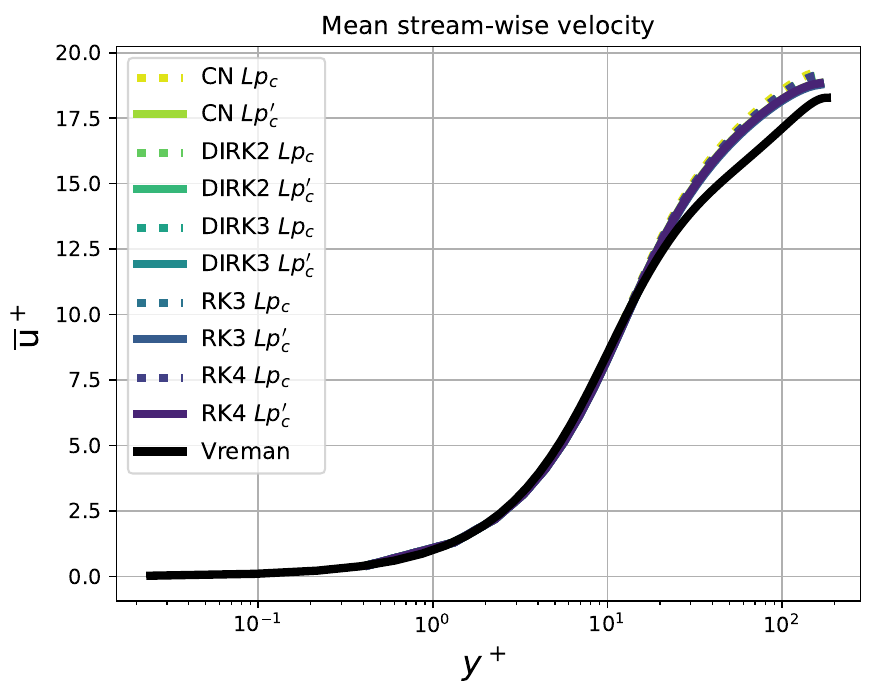}      
	\includegraphics[width=0.32\linewidth]{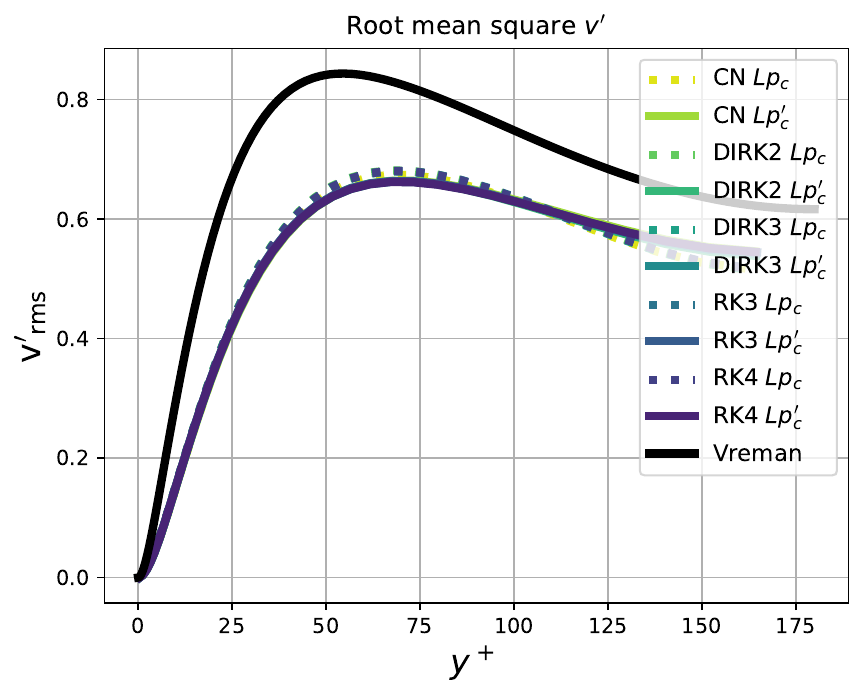}   
	\includegraphics[width=0.32\linewidth]{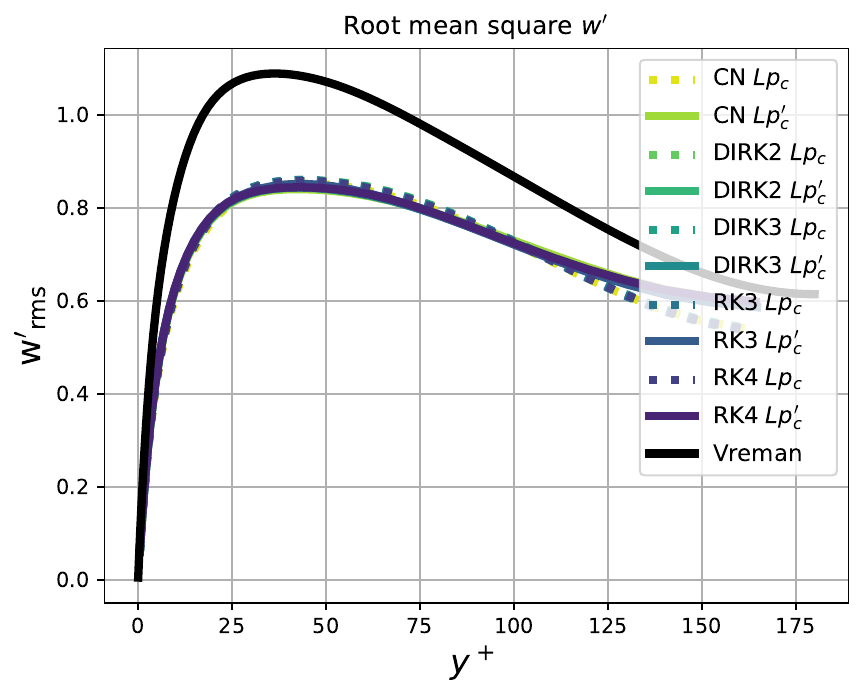}           
	\includegraphics[width=0.32\linewidth]{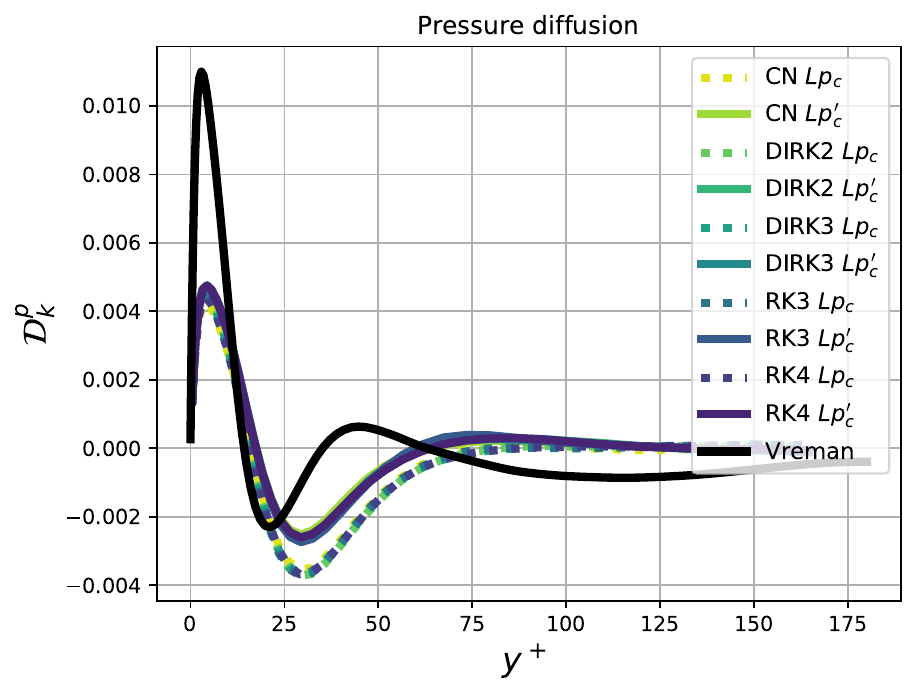}    
	\includegraphics[width=0.32\linewidth]{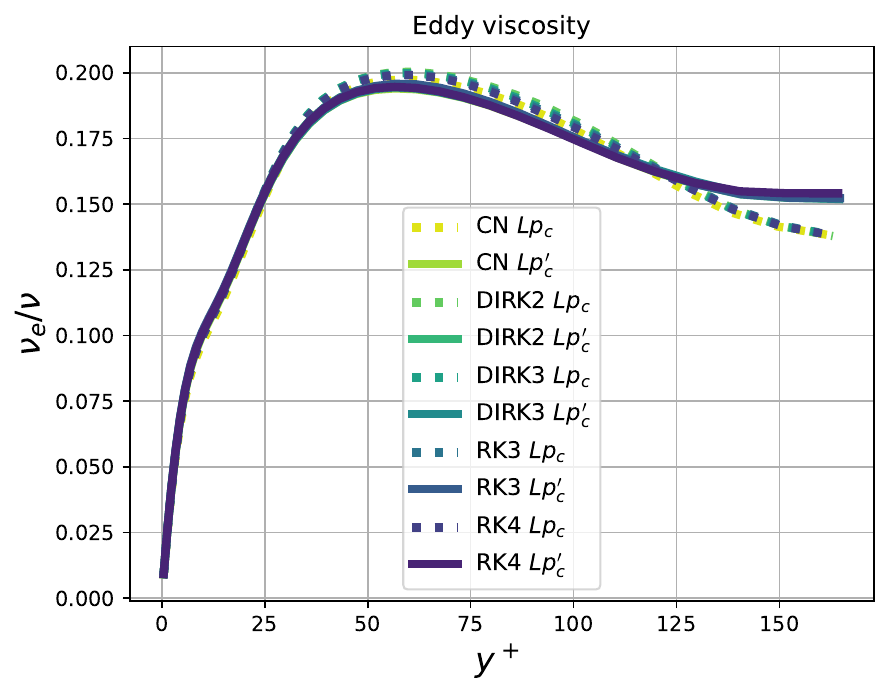}
	\includegraphics[width=0.32\linewidth]{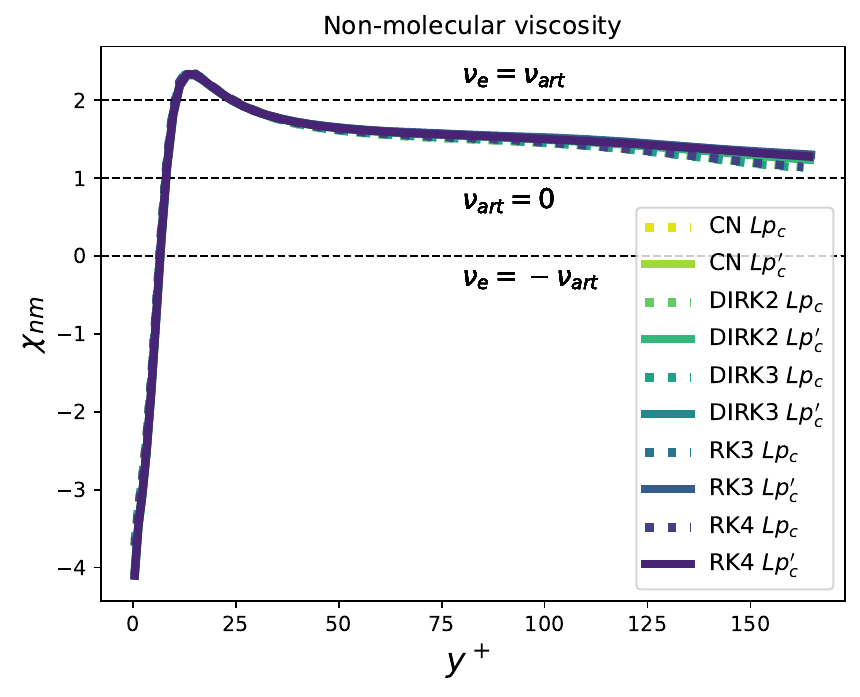} 
	\captionsetup{font = {footnotesize}}
	\caption{Turbulent flow variables in second-, third-, and fourth-order time discretization schemes in channel flow simulations at $\mathrm{Re}_\tau = 180$.}
	\label{fig:Lpc_time}
\end{figure}
In this section, we compare two pressure correction approaches, specifically the first-order Chorin method (denoted as  $Lp_c$) and the second-order Van Kan method (denoted as $Lp_c'$), for addressing the checkerboard problem caused by the velocity-pressure decoupling on collocated grids. The employed six temporal schemes and four model coefficients are listed in Table \ref{tab:ddt1}. A dynamic time-step is employed to ensure a maximum CFL number of 0.4. Here, the symmetry-preserving discretization with the QR model (C=0.101) is applied on a $48\times 76\times48$ mesh at $\mathrm{Re}_\tau=180$.

The effect of pressure correction methods is most significant in the low-order accurate schemes, particularly the Backward-Euler method.  Figure \ref{fig:pn_1st} focuses on the mean streamwise velocity $\bar u$, Reynolds stress, turbulent kinetic energy $k$ and budget terms obtained from the first-order schemes. 
For the Backward-Euler method, the Chorin approach fails to provide accurate predictions. The artificial dissipation is nearly zero, because all budget terms are nearly zero. In contrast, the Van Kan method significantly improves the accuracy of the mean streamwise velocity, Reynolds shear stress, and turbulent kinetic energy ($k$).  
For the Forward-Euler scheme, all results deviate substantially from the DNS reference, and pressure correction methods show negligible influence on most flow variables. However, for quantities such as turbulent kinetic energy ($k$), spanwise Reynolds stress ($w'w'$), pressure diffusion ($\mathcal D^p_{k}$), and artificial dissipation ($\epsilon^{art}_k$, $\nu_{art}$), the Van Kan method provides more accurate predictions. This indicates that the Van Kan method provides better performance for certain critical parameters even in low-order schemes.

Figure \ref{fig:Lpc_time} compares the performance of the Van Kan and Chorin methods across second-, third-, and fourth-order temporal discretization schemes. The results indicate that the pressure correction has a dominant effect on the accuracy of the prediction. The Chorin method produces consistent results across various temporal orders, and the Van Kan method renders the flow quantities insensitive to the temporal discretization. 

Figure \ref{fig:Lpc_time} shows that in the region $25 < y^+ < 100$, the flow quantities such as $\bar u$, $u'u'$, $v'v'$, $u'v'$, and $k$ are higher for the Chorin method than for the Van Kan method. In the channel center ($y^+ > 125$), the opposite trend is observed, with the Chorin method yielding lower values. For the turbulent kinetic energy budget terms, the most notable difference is seen in the pressure diffusion term ($\mathcal{D}^p_k$). 
The Chorin method predicts a higher negative peak in the region $25 < y^+ < 75$, showing a stronger pressure transport mechanism compared to the Van Kan method. This local pressure transport transfers more turbulent kinetic energy from the streamwise ($u'$) to the wall-normal ($v'$) and spanwise ($w'$) components, leading to improved local predictions of $v'$ and $w'$. 

In terms of eddy viscosity and artificial viscosity, the two methods produce consistent results for $y^+ < 125$. However, in the channel center, the Chorin method introduces lower eddy and artificial viscosities compared to the Van Kan method. 

\subsubsection{Impact of Spatial Discretizaiton}
\begin{figure}[!b]
	\centering  
	\includegraphics[width=0.32\linewidth]{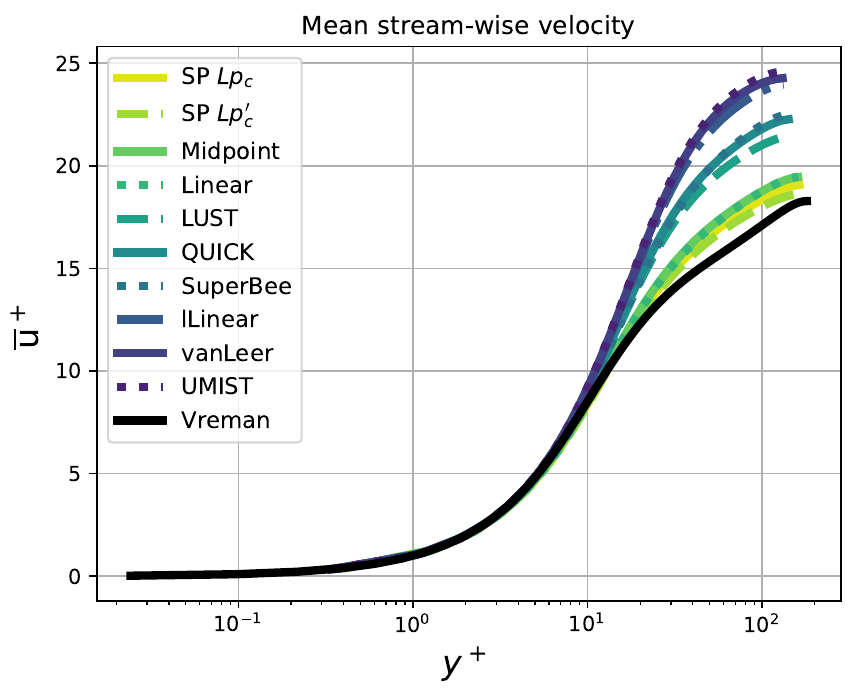} 
	\includegraphics[width=0.32\linewidth]{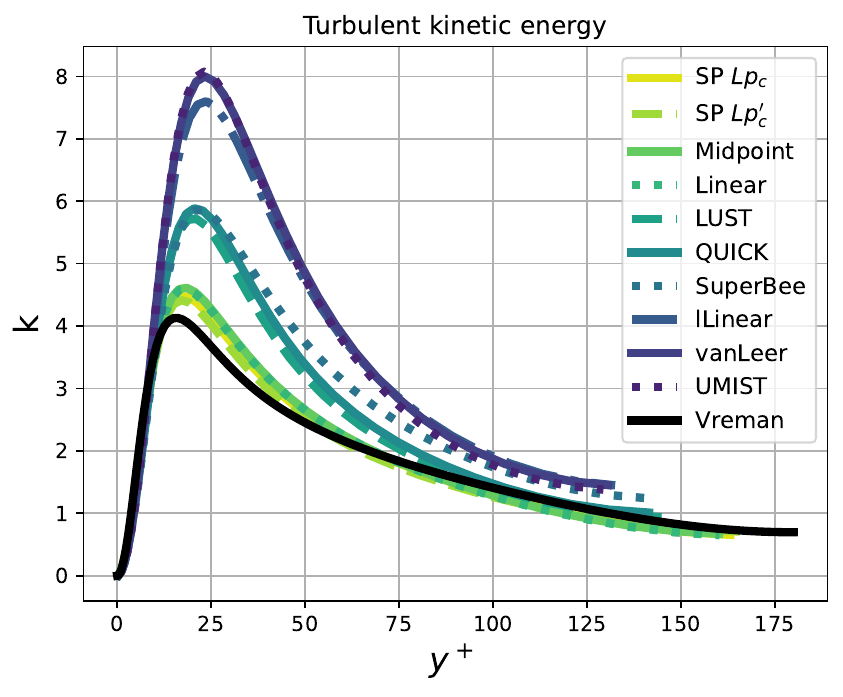} 
	\includegraphics[width=0.32\linewidth]{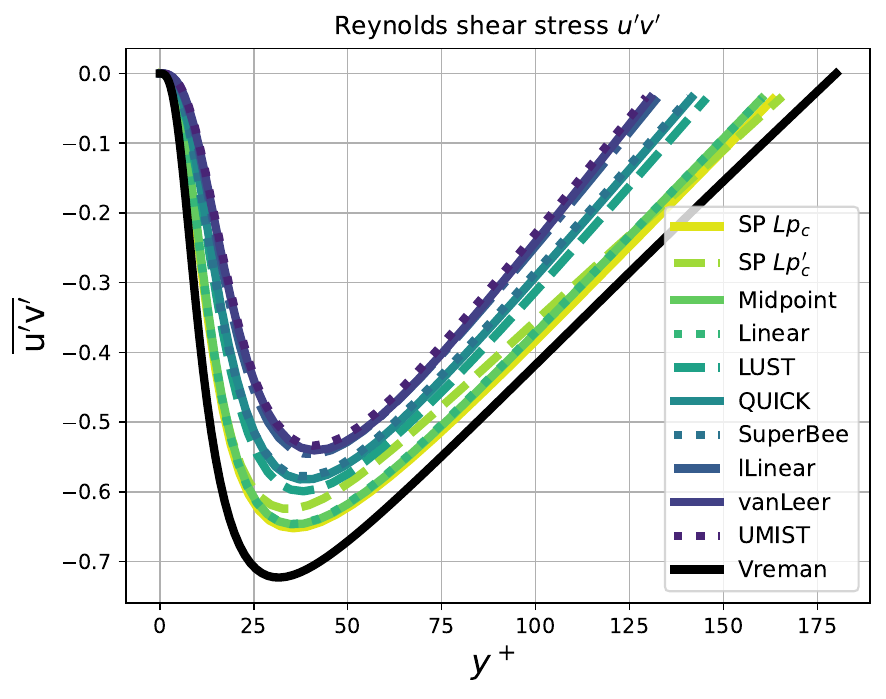} 
	\includegraphics[width=0.32\linewidth]{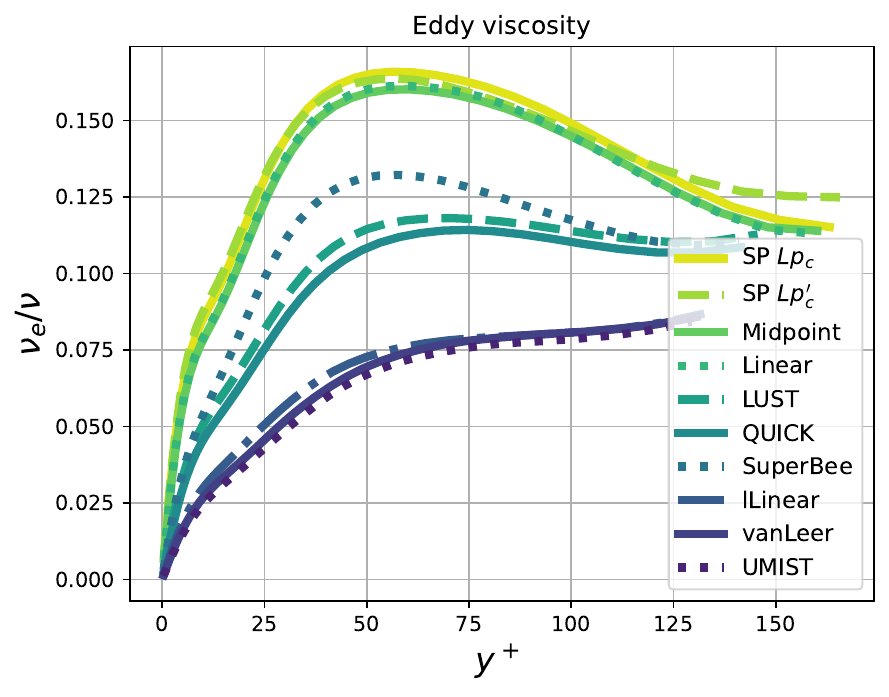} 
	\includegraphics[width=0.32\linewidth]{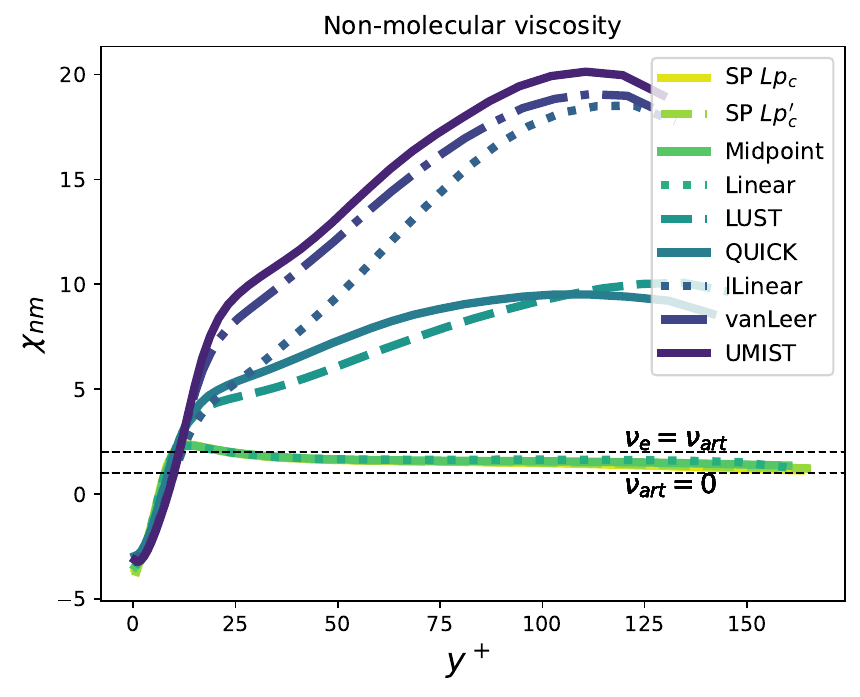} 
	\includegraphics[width=0.32\linewidth]{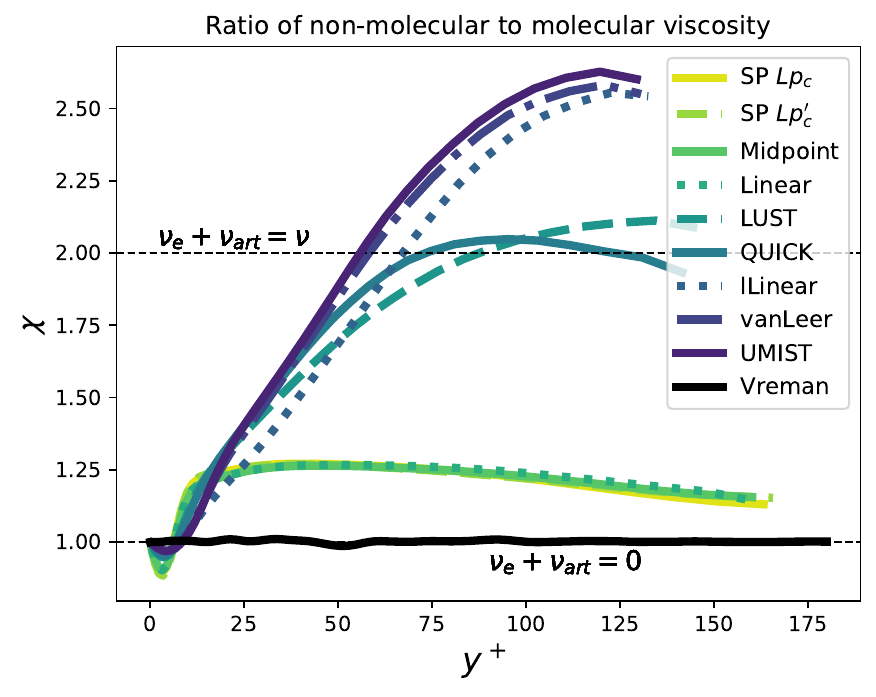}  
	\captionsetup{font = {footnotesize}}
	\caption{Flow variables in channel flow simulations at $\mathrm{Re}_\tau = 180$, employing various divergence schemes with the QR model coefficients $C = 0.083$. lLinear is limited-Linear with a coefficient 0.2. $Lp_c$ and $Lp_c'$ refer to the Chorin and van Kan method, respectively.}
	\label{fig:div}
\end{figure}
The present investigation evaluates various spatial discretization schemes in OpenFOAM, including Upwind, Mid-point, Linear, vanLeer, limitedLinear (with a scalar bound of 0.2), SuperBee, LUST (Linear-Upwind Stabilised Transport), QUICK (Quadratic Upstream Interpolation for Convective Kinematics), UMIST (Upstream Monotonic Interpolation for Scalar Transport), and symmetry-preserving discretization conjugated with the Chorin and van Kan pressure corrections \cite{komen2021symmetry}. The PIMPLE algorithm is employed with standard OpenFOAM schemes, while the symmetry-preserving discretization is implemented using the RKSymFOAM solver \cite{janneshopmanRKSymFoam}. A QR model coefficient of $C=0.083$ is applied (the coefficients of 0.083 and 0.101 give almost equal predictions), with a mesh resolution of $48 \times 76 \times 48$ grid points. Temporal discretization is performed using the second-order accurate Crank-Nicolson scheme. The results, presented in Figure \ref{fig:div}, exclude the Upwind scheme due to its failure to accurately predict any flow quantities.

Figure \ref{fig:div} shows the mean-streamwise velocity, turbulent kinetic energy and Reynolds stress. The results can be categorized into three groups.
The most accurate predictions are achieved with the symmetry-preserving, Mid-point, and Linear schemes, with the symmetry-preserving approach showing a slight advantage over the other two. The Mid-point and Linear schemes produce nearly identical results across all flow variables. The next group consists of the LUST, QUICK, and SuperBee schemes, while the least accurate predictions are provided by the third group consists of the limitedLinear, vanLeer, and UMIST schemes.
 
Although the SuperBee scheme provides reasonable accuracy for the streamwise components, it exhibits notable deviations in the spanwise and wall-normal components. These deviations are attributed to discrepancies in the pressure diffusion mechanism, which redistributes energy from the streamwise direction to the other two. The pressure diffusion predicted by SuperBee does not vanish in the bulk region; instead, it continues to transfer energy to the $v'v'$ and $w'w'$ components. Additionally, the scheme underestimates viscous dissipation, even in the channel center where other schemes exhibit self-similarity. Moreover, the artificial viscosity predicted by SuperBee is largely negative, artificially generating turbulent kinetic energy and destabilizing the simulation. 

For eddy viscosity $\nu_e$ and artificial viscosity $\nu_{art}$, all the schemes overlap in the log-wall region ($y^+<10$), with discrepancies mainly appearing in the outer region $y^+>10$. The most accurate schemes produce the highest eddy viscosity (approximately $15\% \nu$) and the lowest artificial viscosity ($0 < \nu_{art} < \nu_e$), with a non-molecular viscosity ratio in the range $1 <\chi_{nm}< 2$. In contrast, for the other two groups of schemes (with SuperBee excluded due to large negative values), the eddy viscosity is lower ($\nu_e < 12\% \nu$), and the artificial viscosity is significantly higher than the eddy viscosity ($5 <\chi_{nm}< 20$), resulting in a total non-molecular viscosity exceeding the molecular viscosity ($\chi > 2$). Consequently, the LES model is severely suppressed when spatial discretization schemes introduce excessive numerical error, such that $\nu_{art}$ $\mathtt{\sim}$ $\nu$.  

In conclusion, the symmetry-preserving discretization outperforms other standard OpenFOAM divergence schemes in this case study, demonstrating superior accuracy in the prediction of key flow quantities.

\paragraph{Mesh Convergence Study}
\label{sec:QRmesh}
\begin{table}[b!] 
\footnotesize
\centering
\begin{tabular}{cccccccccccccc}
\toprule
Case & $N_x$ & $N_y$ & $N_z$ & $\Delta x^+$ & $\Delta y^+_w$ & $\Delta y^+_b$ & $\Delta z^+$ & $N_{tot}$\\
\midrule 
M0&26&78&26&90&0.88&13&30& 52,728\\
M1&38&78&38&60&0.88&13&20&112,632\\
M2&48&76&48&47 &0.9 &14 &16 &175,104 \\
M3&64&78&76&35&0.88&13&10&379,392\\
M4&128&78&76&18&0.88&13&10&758,784\\
M5&128&78&128&18&0.88&13&6&1,277,952 \\
M6&256&176&168&9&0.39&5.9&4.5&7,569,408\\
Vreman&384&193&192&5.9&0.0024&-&3.9&14,229,504\\
\bottomrule
\end{tabular} 
\captionsetup{font={footnotesize}}
\caption{Mesh convergence study using symmetry-preserving discretization with QR model coefficients C=0.012, 0.048, 0.083, 0.092, 0.101, and 0.125. Stretching ratio $SR=15$. $N_{tot}$ denotes total mesh points.}
\label{tab:QRmesh1}
\end{table} 
\begin{figure}[!b]
	\centering 
	\includegraphics[width=0.32\linewidth]{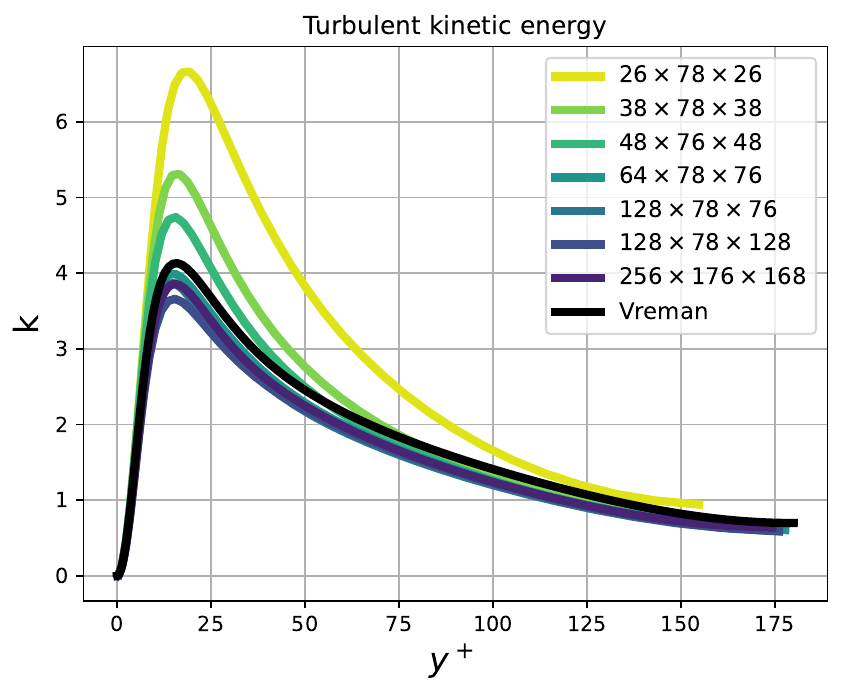}
	\includegraphics[width=0.32\linewidth]{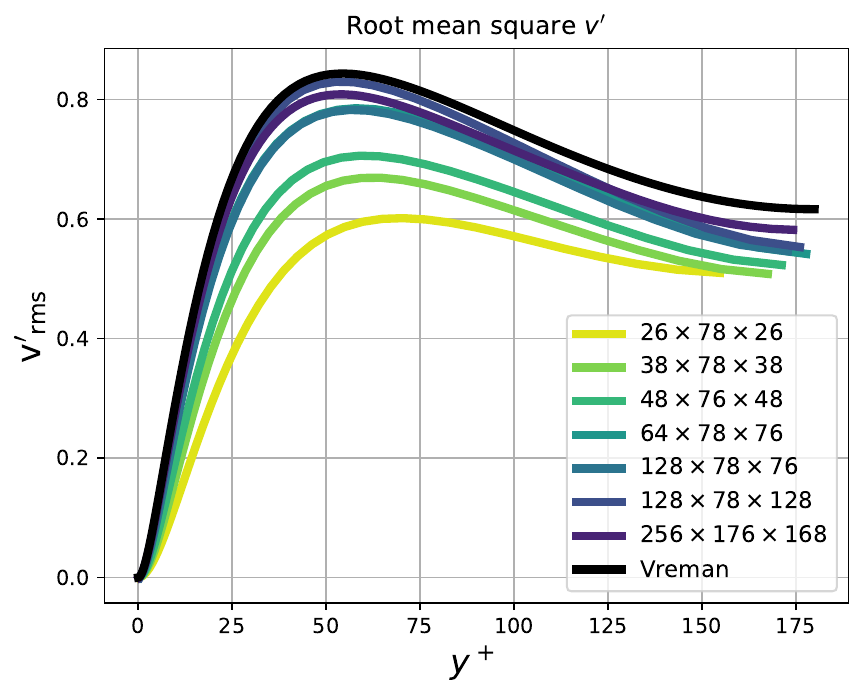}   
	\includegraphics[width=0.32\linewidth]{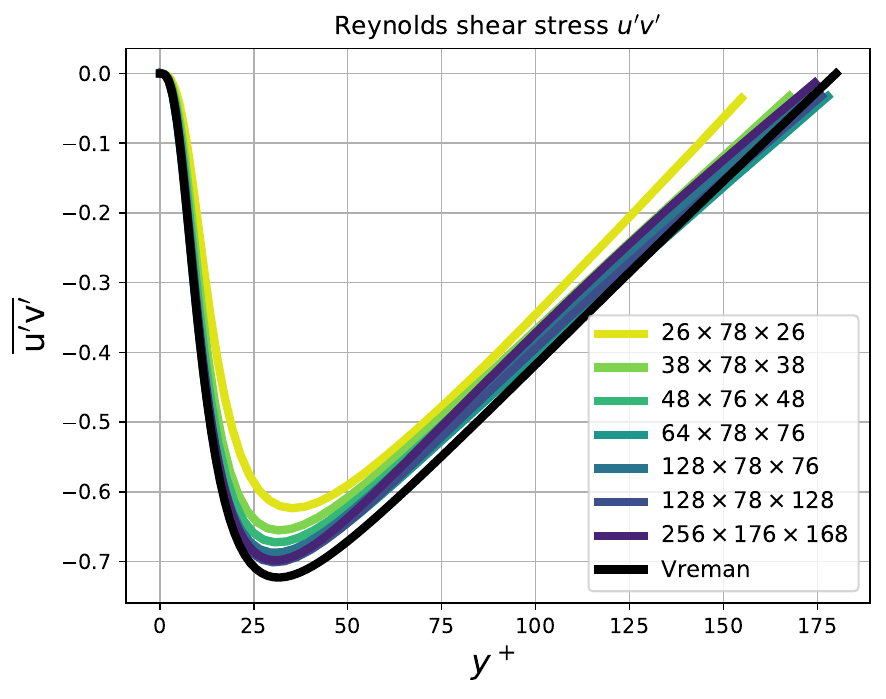} 
	\includegraphics[width=0.32\linewidth]{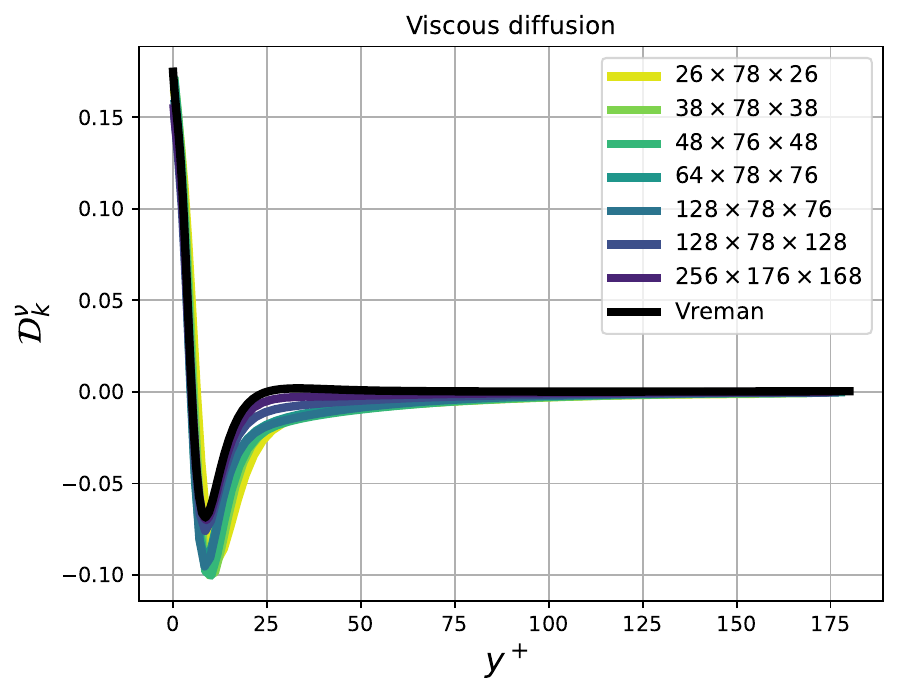}  
	\includegraphics[width=0.32\linewidth]{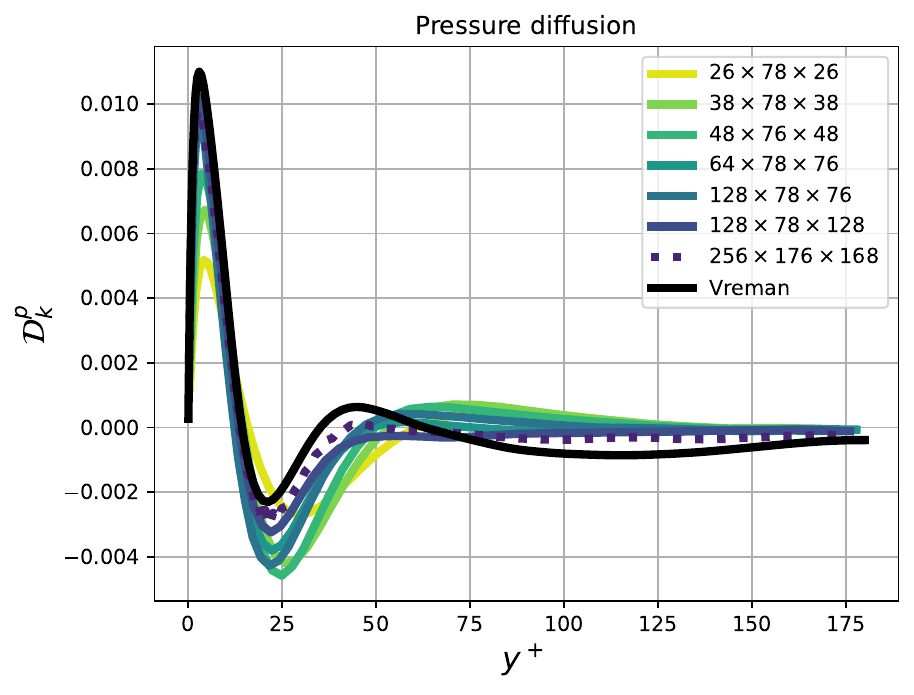}  
	\includegraphics[width=0.32\linewidth]{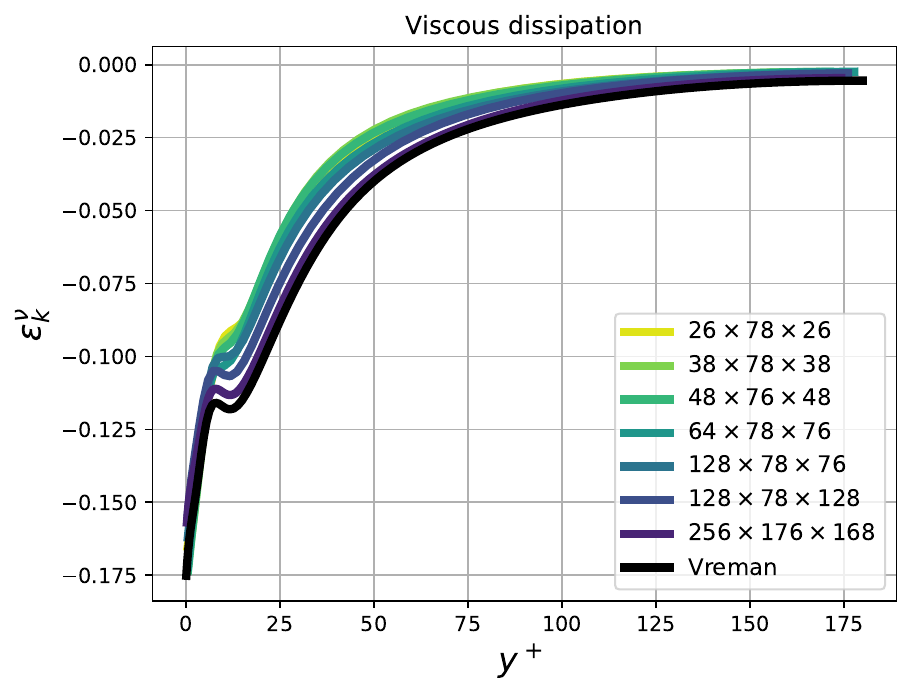}
	\includegraphics[width=0.32\linewidth]{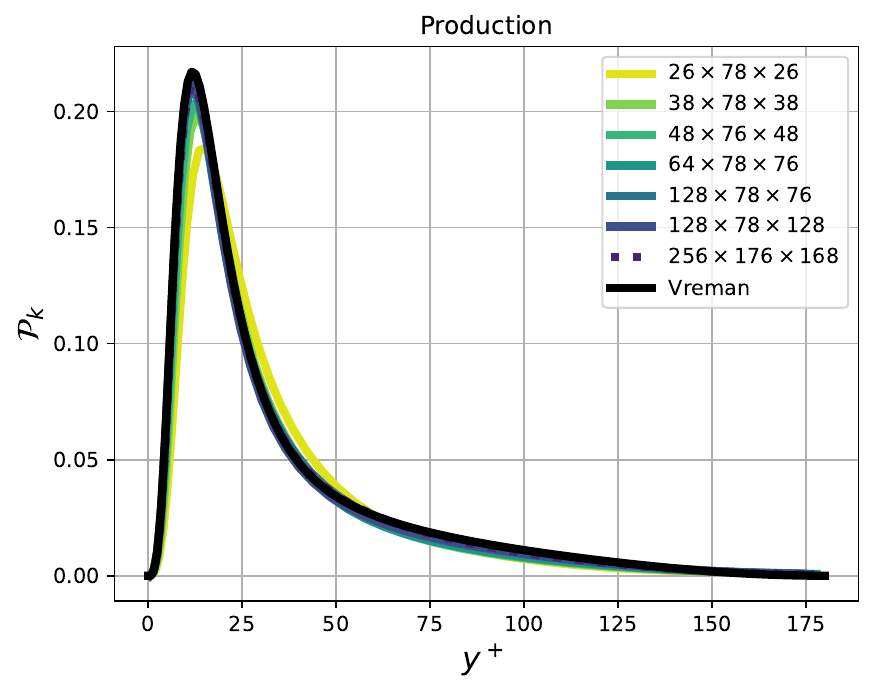}   
	\includegraphics[width=0.32\linewidth]{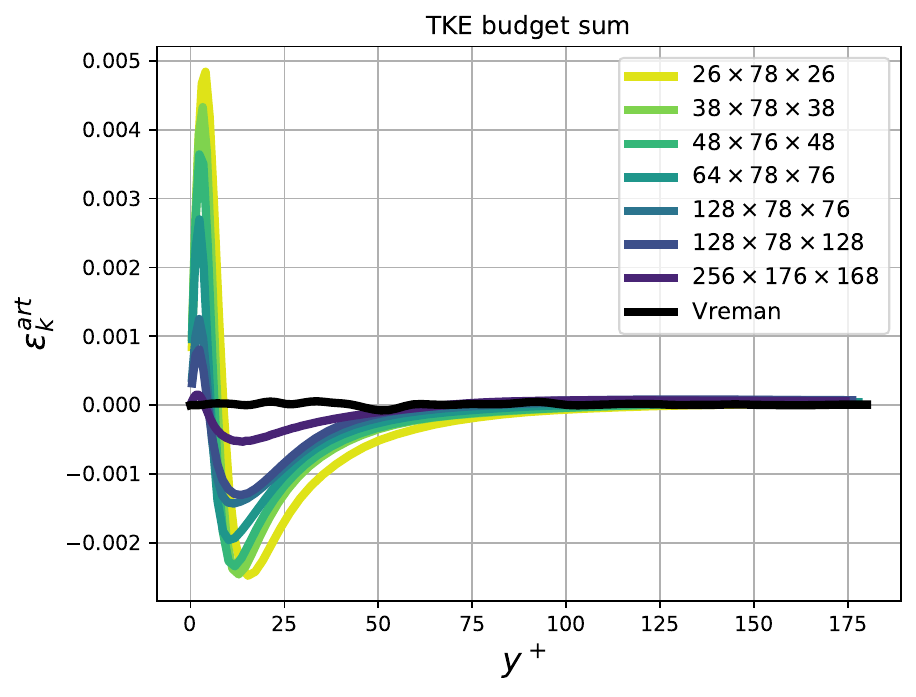}
	\includegraphics[width=0.32\linewidth]{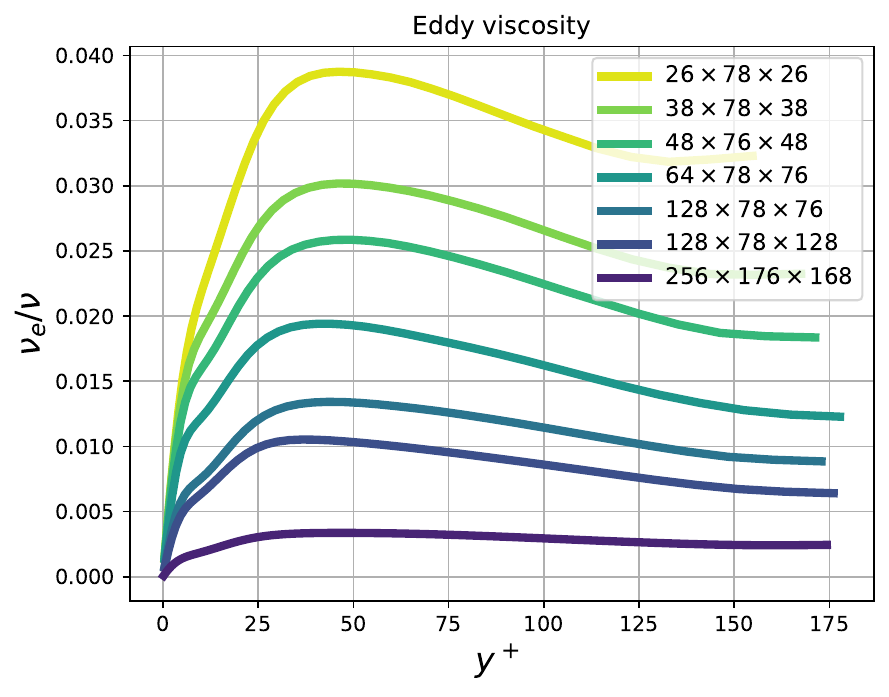}   
	\captionsetup{font = {footnotesize}}
	\caption{Mean, turbulent kinetic energy, energy balance terms and eddy viscosity  in channel flow simulations at $\mathrm{Re}_\tau=180$ for mesh convergence study using symmetry-preserving discretization and the QR model with $C=0.012$.}
	\label{fig:QRmesh1}
\end{figure}
%\begin{figure}[!h]
%	\centering
%%	\includegraphics[width=0.32\linewidth]{180_QR101_M1_Q0_01.pdf}   
%	\includegraphics[width=0.48\linewidth]{180_QR101_M4_Q0_01_bottom.pdf}
%	\includegraphics[width=0.48\linewidth]{180_QR012_M5_Q200.pdf} 
%	\captionsetup{font = {footnotesize}}
%%	\caption{Q-criteria on mesh configuration of M1 (left), M5(middle) and M6(right). }
%	\caption{Q-criteria on mesh configuration of M5 (left) and M6(right). }
%	\label{fig:contour}
%\end{figure}
 
The mesh configurations listed in Table \ref{tab:QRmesh1} are based on the mesh guideline for the wall-resolved LES  proposed by Georgiadis et al. \cite{georgiadis2010}, namely,
\begin{equation}
50\leq\Delta x^+\leq 150, \quad \Delta y^+ <1,  \quad 15\leq\Delta z^+\leq 40. \nonumber
\end{equation}  

The flow variables are illustrated in Figure \ref{fig:QRmesh1}. The profiles of the mean streamwise velocity $\bar u^+$, RMS of $u'$ and $w'$, turbulent kinetic energy $k$, and budget terms $\mathcal D^{\nu}_k$ and $\epsilon^{\nu}_k$ show monotonic convergence to DNS results with mesh refinement.
The eddy-related terms ($\mathcal D^{sgs}_k, \epsilon^{sgs}_k, \mathcal P_{sgs}$, and $\nu_e$) and artificial error terms ($\epsilon^{art}_k$ and $\nu_{art}$) decrease monotonically with mesh refinement. However, non-monotonic behavior is observed in the flow quantities $v'_{rms}, u'v'$, and $\mathcal D^{p}_k$. Accuracy degrades on the finest mesh M6 ($256\times176\times168$) compared to the second finest mesh M5 ($128\times78\times128$).  The viscous dissipation rate $\epsilon^{\nu}_k$, which reflects activities at the smallest scales of turbulence, requires very fine meshes for accurate capture. On the finest mesh (M6) with $C=0.012$, dissipation rates exhibit closer convergence to DNS data compared to $C=0.048, 0.083, 0.101$, and $0.125$.  Within the current mesh configurations, eddy viscosity remains below $\frac{\nu_e}{\nu} < 5\%$ and exhibits a nearly uniform distribution in the wall-normal direction. 

The explanations of the flow patterns in Figure \ref{fig:QRmesh1} are given here.   Comparing the artificial dissipation $\epsilon^{art}_k$ on the two finest meshes (M5 and M6) in Figure \ref{fig:QRmesh1}. $\epsilon^{art}_k$ is almost all negative on the finest mesh M6, acting as a sink term, similar to $\epsilon^\nu_k$ and $\epsilon^{sgs}_k$. This results in more dissipation in $k$ and $u'$ compared to M5, explaining the monotonic convergence of $k$ and $u'$.\\
In contrast, pressure diffusion $\mathcal D^p_k$, which redistributes turbulent kinetic energy from the streamwise to the spanwise and wall-normal directions, is less vigorous on M6 and deviates more from DNS data compared to M5 in the near wall region $y^+<20$, leading to less transfer of turbulent kinetic energy from the streamwise to the spanwise direction in the near wall region. This reduced accuracy in $\mathcal D^p_k$ on the finer mesh explains the degradation in $v'_{rms}$ accuracy.  \\
For the shear stress $u'v'$, the underprediction likely results from a slight underestimation of $\mathcal P_k$ on M6 compared to M5. Since the mean flow is correctly captured, the underprediction in $\mathcal P_k = \overline{u'v'}\hspace{0.2em}\overline{\partial_j u_i}$  can only be reflect by the shear stress $u'v'$, explaining the non-monotonic behavior observed in $u'v'$.  \\
The monotonic reduction in eddy-related terms is attributed to two intertwined mechanisms. 
First, increasing the number of grid points provides more cell data for the discretization and interpolation of flow quantities, reducing spatial-truncation error.  Second, as more scales are resolved ($\Delta\rightarrow 0$), the contribution of the sub-grid-scale model diminishes. 

\paragraph{Subgrid-Scale Model Coefficient and Mesh Resolution}
\begin{figure}[!b]
	\centering
	\includegraphics[width=0.45\linewidth]{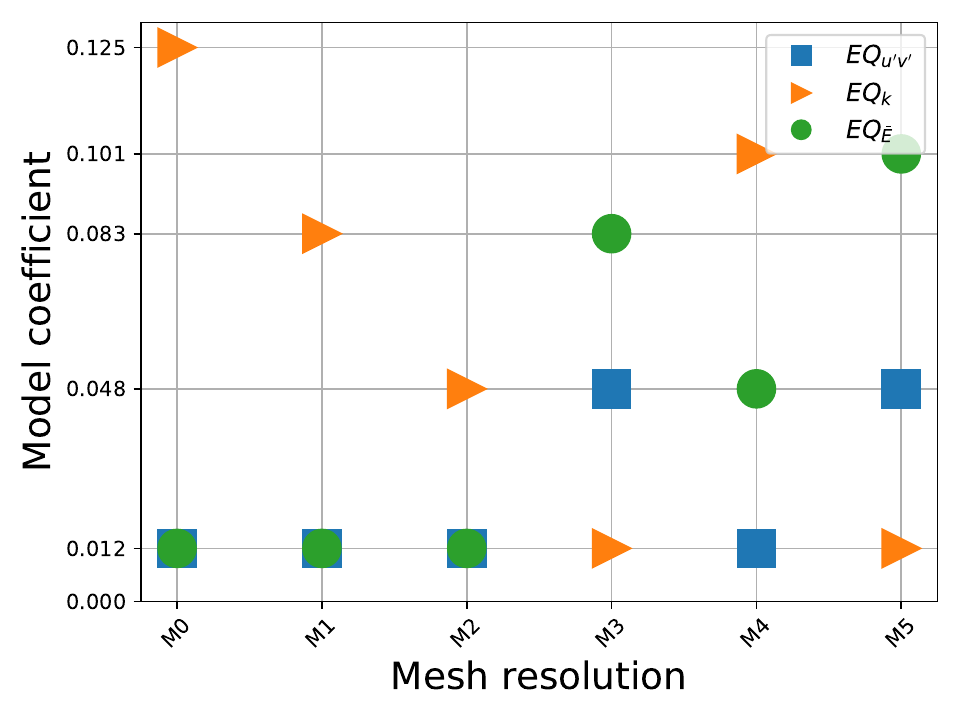} 
	\captionsetup{font = {footnotesize}}
	\caption{Optimal QR model coefficient determined through error quantification for the Reynolds shear stress $u'v'$, turbulent kinetic energy $k$, and kinetic energy of the mean flow $\bar E$, plotted against mesh resolution. The x-axis labels ($M0$ through $M5$) correspond to the mesh resolutions listed in Table \ref{tab:QRmesh1}. Symbols ($\square, \triangleright$, and $\circ$) represent the error quantification results for $u'v'$, $k$, and $\bar E$ respectively.}
	\label{fig:QRmesh2}
\end{figure} 
The dependence of the subgrid model coefficient on the mesh size has been reported in various studies, such as Meyers et al. \cite{MEYERS2007156} and Lucor et al. \cite{LUCOR2007}, focusing on the classic Smagorinsky model coefficient in decaying homogeneous isotropic turbulence. 
They observed that the Smagorinsky model coefficient's dependency on resolution is non-monotonic and can be irregular for marginal simulation resolutions. 
 
In this study, we investigate the mesh dependence of the QR model coefficient. As shown in the second part of the present series of articles,  increasing the model coefficient beyond a certain threshold can lead to turbulence laminarization, resulting in smoother velocity fields and reduced turbulent kinetic energy transfer from the streamwise to spanwise and wall-normal directions. 
Error quantification has revealed that the error in $u'v'$ increases monotonically after reaching the optimal model coefficient. Consequently, we narrowed the coefficients to $C=0.012, 0.048, 0.083, 0.101,$ and $0.125$ for investigating the mesh dependence. 
Across these five coefficients, six distinct mesh configurations in Table \ref{tab:QRmesh1} were applied. The results are presented in Figure \ref{fig:QRmesh2}. 

Figure \ref{fig:QRmesh2} shows the optimal QR coefficient for the Reynolds shear stress $u'v'$, turbulent kinetic energy $k$, and kinetic energy of the mean flow $\bar E$. 
From Figure \ref{fig:QRmesh2}, 
 it is evident that on the three coarsest meshes, a model coefficient of $C=0.012$ provides the most accurate predictions for both $\bar{E}$ and $u'v'$, while for turbulent kinetic energy ($k$), the optimal coefficient varies: $C=0.125$ for M0, $0.083$ for M1, and $0.048$ for M2. 
 As the mesh is refined, larger coefficients yield better predictions for $\bar{E}$, specifically $C=0.083$ for M3, $0.048$ for M4, and $0.101$ for M5. Conversely, smaller coefficients are more effective for capturing $k$. Despite these trends, no consistent optimal coefficient has emerged across all meshes.

The general patterns for optimal predictions of $u'v'$, $k$, and $\bar{E}$ can be summarized as follows: For Reynolds shear stress $u'v'$, the most accurate results are obtained when non-molecular dissipation is minimal, favoring smaller model coefficients. For turbulent kinetic energy $k$, which represents small-scale motions, coarser meshes require larger model coefficients (and thus greater eddy dissipation) for accurate predictions, while finer meshes necessitate less eddy damping. In contrast, for the energy of the mean flow $\bar{E}$, representing large-scale motions, relatively smaller coefficients on coarse mesh produce the most accurate results while finer meshes necessitate more eddy damping.

\paragraph{Molecular Viscosity vs. non-molecular Viscosity}
\begin{figure}[!h]
	\centering  
	\includegraphics[width=0.32\linewidth]{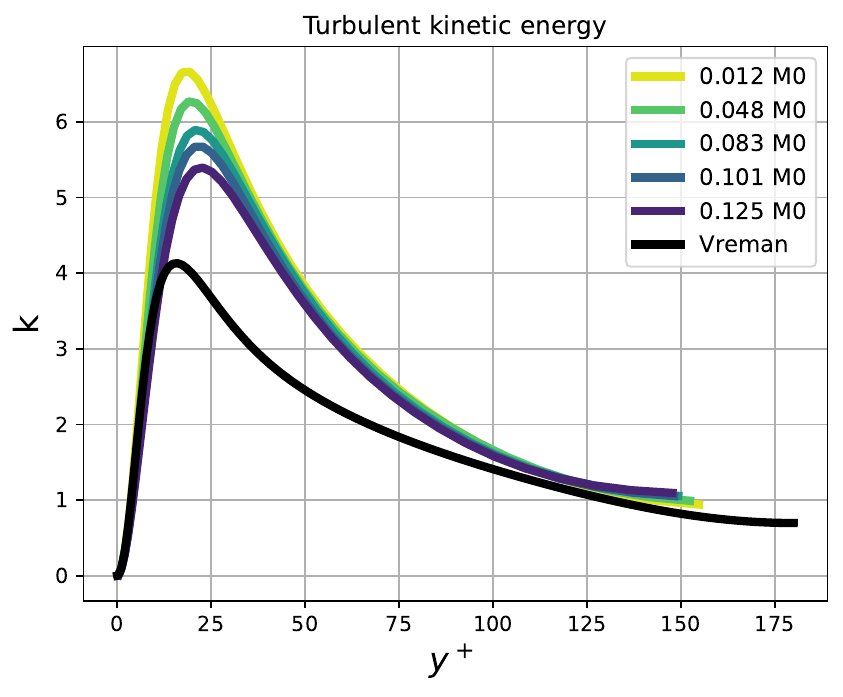}
	\includegraphics[width=0.32\linewidth]{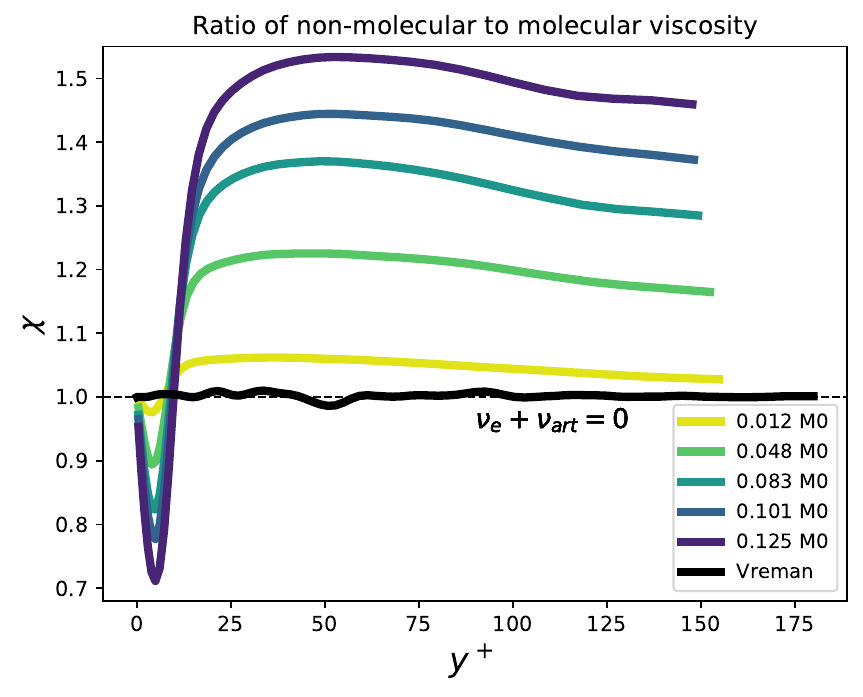}
	\includegraphics[width=0.32\linewidth]{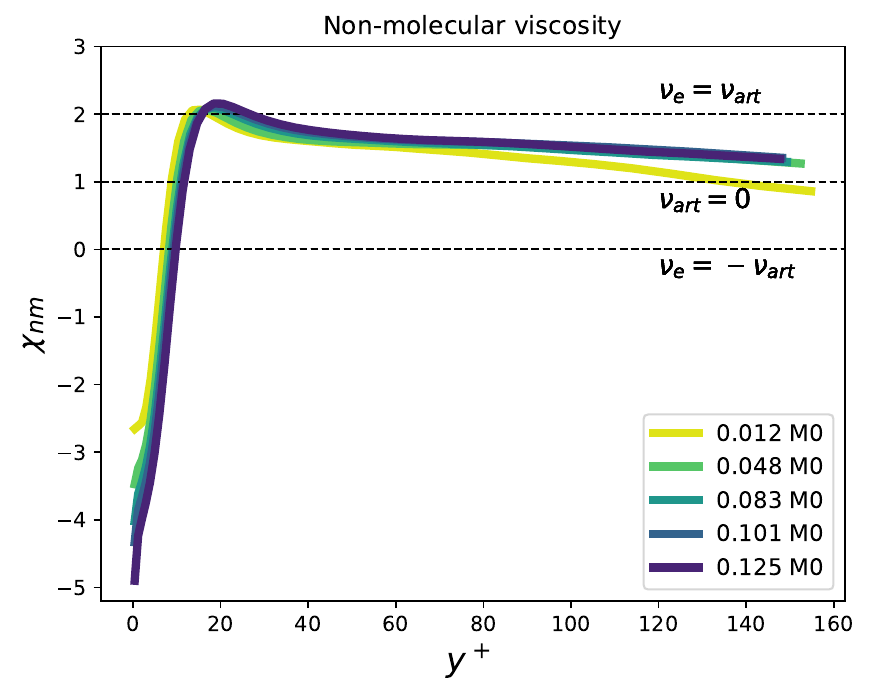}
	\includegraphics[width=0.32\linewidth]{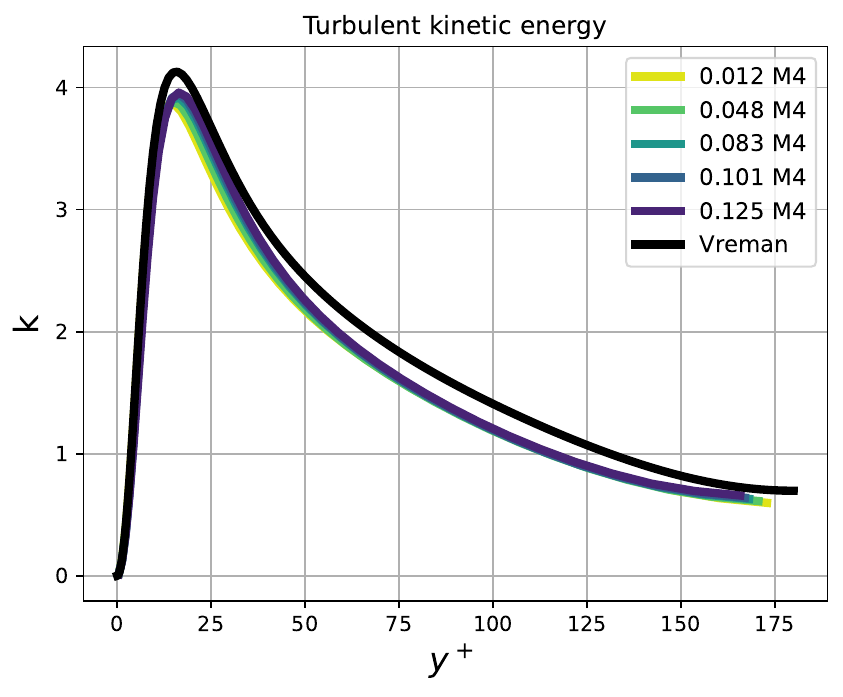}
	\includegraphics[width=0.32\linewidth]{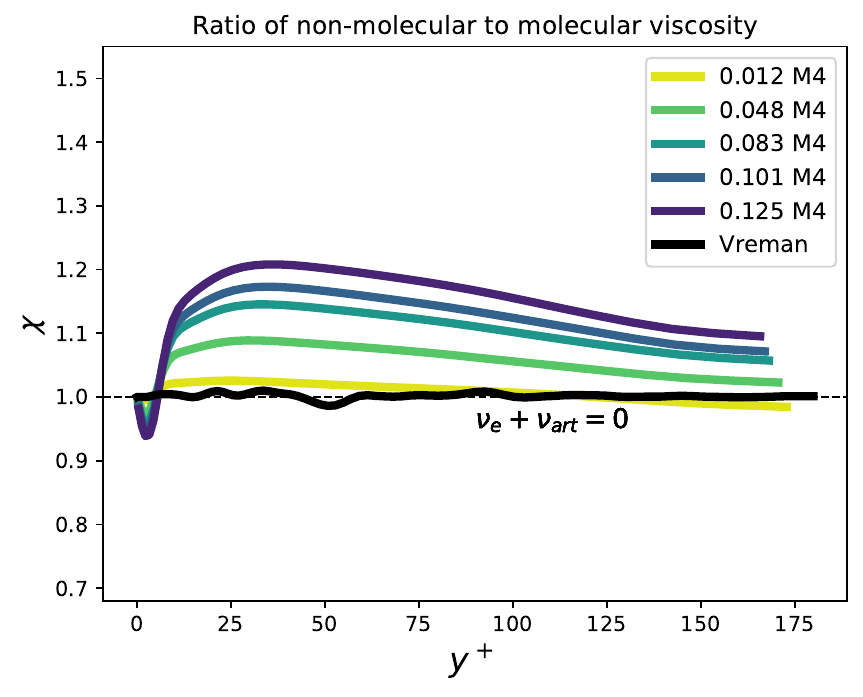}
	\includegraphics[width=0.32\linewidth]{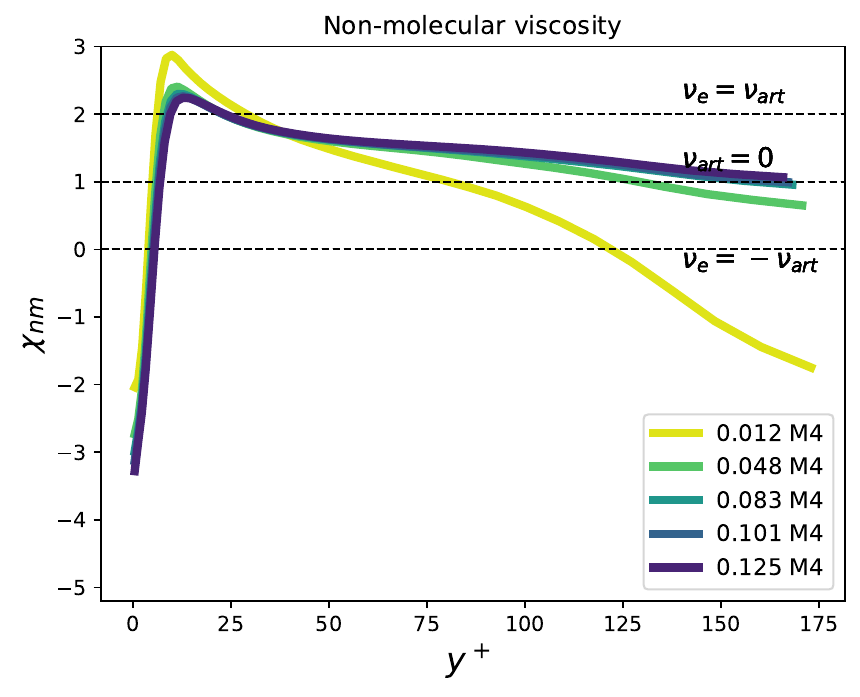}
	\includegraphics[width=0.32\linewidth]{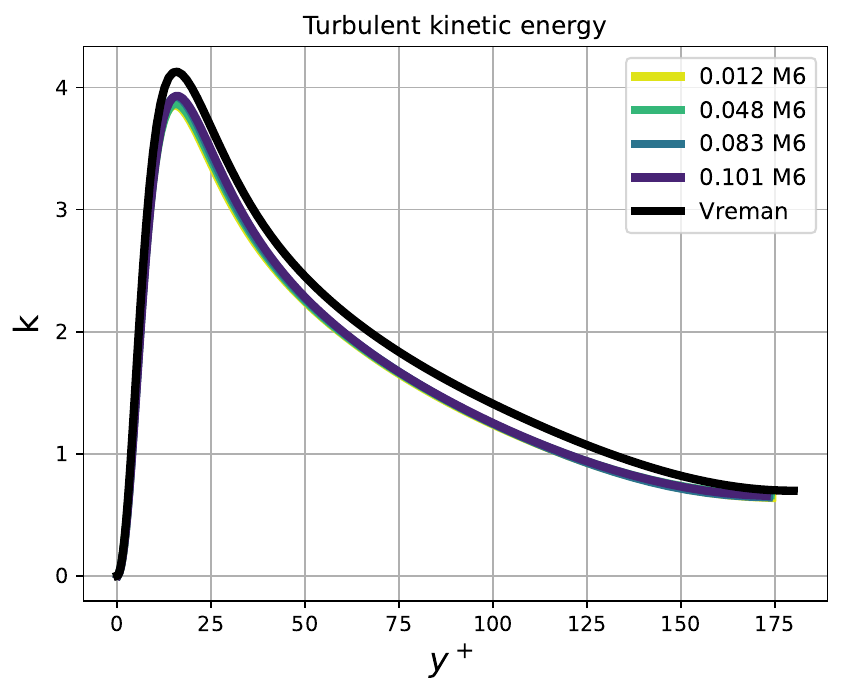}
	\includegraphics[width=0.32\linewidth]{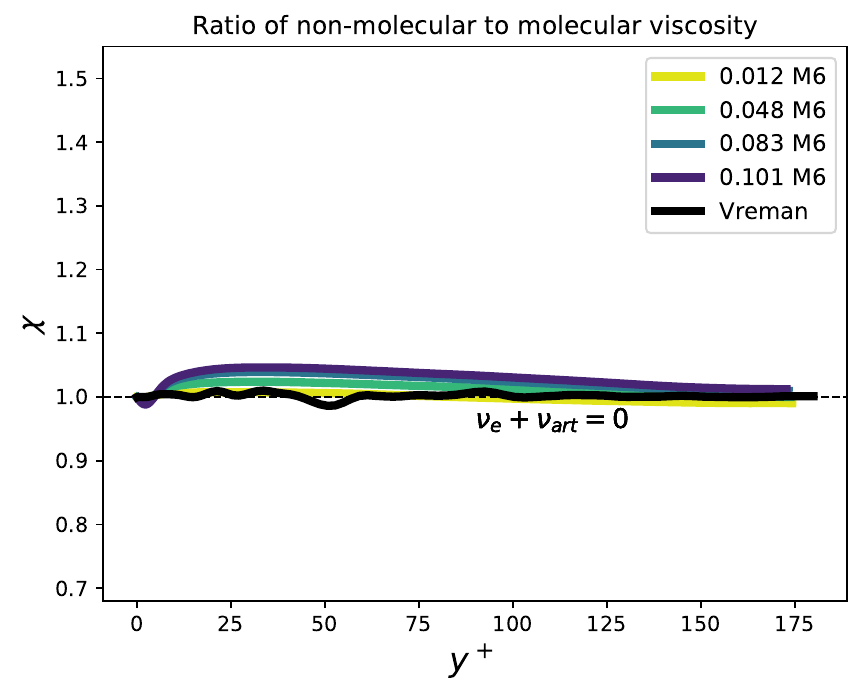}
	\includegraphics[width=0.32\linewidth]{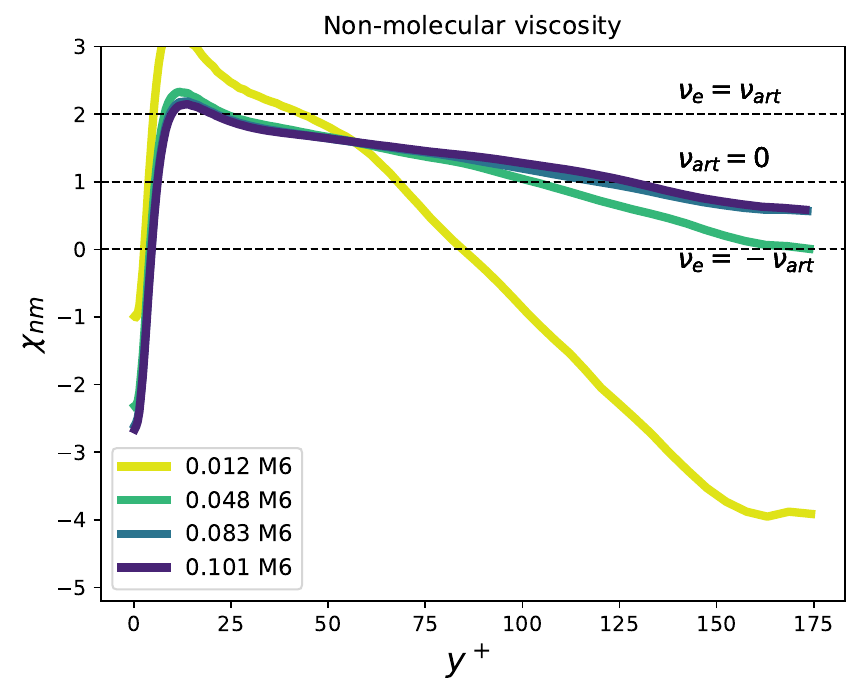}  
	\captionsetup{font = {footnotesize}}
	\caption{Eddy viscosity and artificial viscosity  in channel flow simulations at $\mathrm{Re}_\tau=180$ on different meshes using symmetry-preserving discretization and the QR model with $C=0.012, 0.048, 0.083, 0.101$ and 0.125. The y-axis scale for $\chi$ and $\chi_{nm}$ is unified across all meshes for comparison. }
	\label{fig:QRmesh3}
\end{figure} 
Figure \ref{fig:QRmesh3} illustrates the turbulent kinetic energy ($k$) and the viscous parameters $\chi$ and $\chi_{nm}$ across different mesh resolutions.
For the non-molecular viscous parameter $\chi_{nm}$, within the near-wall region ($y^+ < 25$), $\chi_{nm}$ increases monotonically from a negative maximum to a positive peak. When $\chi_{nm} < 1$, the artificial viscosity $\nu_{art}$ is negative, resulting in the production of turbulent kinetic energy ($k$).
As $\chi_{nm}$ approaches zero, the relation $\nu_{art} \approx -\nu_e$ holds, and further decreases to $\chi_{nm} \approx -5$ indicate that $\nu_{art} \approx -6\nu_e$. In this case, the eddy viscosity $\nu_e$ dampens $k$, while $\nu_{art}$ generates it, with the two opposing each other. This pronounced negative value of $\chi_{nm}$ stems from the QR model switching off at the no-slip wall, leading to a rapid decline in eddy viscosity near the wall.
In the region $y^+ > 25$, the range $0< \chi_{nm} <2$ suggests that $-\nu_e < \nu_{art} <\nu_e$. For model coefficients $C \geq 0.048$, $\chi_{nm}$ becomes insensitive to changes in SGS model, implying that both eddy viscosity and artificial viscosity vary in a similar manner. The amount of eddy viscosity has a dominant effect on the amount of artificial viscosity, the larger the SGS model coefficient the more eddy viscosity, and more artificial viscosity.

For the ratio of the non-molecular to physical viscous parameter $\chi$, within the near-wall region ($y^+ < 15$), $\chi \leq 1$ indicates that the combined non-molecular viscosity $(\nu_e + \nu_{art} )\leq 0$. Since eddy viscosity is strictly positive ($\nu_e > 0$), this suggests that $\nu_e$ and $\nu_{art}$ counterbalance each other, with the net effect being the production of turbulent kinetic energy ($k$). It is observed that increasing eddy viscosity enhances numerical production in this near-wall region $y^+<15$.
In the outer region ($y^+ > 15$), when $\chi > 1$, the net effect of the non-molecular viscosity dissipates turbulent kinetic energy.
As eddy viscosity increases, artificial dissipation is enhanced, reducing the artificial production of $k$. 
 Only on the coarsest mesh with a large coefficient ($C = 0.125$) does the non-molecular viscosity exceed half of the molecular viscosity. In all other cases, molecular viscosity dominates dissipation compared to the net non-molecular viscosity.

\paragraph{Remarks on Irregular Convergence}
The above mesh convergence study encounters irregular convergence, which commonly occurs in LES. This is a consequence of the nonlinear interaction between the error sources (subgrid model and discretization) that are difficult to explain. We will do so below. Firstly, the accuracy sometimes degrades with mesh refinement.  Particularly, $v'_{rms}, u'v'$, and $\mathcal D^{p}_k$ exhibit non-monotonic convergence. Accuracy degrades on the finest mesh M6 compared to the second finest mesh M5. Secondly, the optimal model coefficient is not mesh independent, but behaves irregular upon mesh refinement.This was also found by Meyers et al. \cite{MEYERS2007156}. To understand these behaviors, we examine the variation of turbulent kinetic energy $k$ and artificial viscosity $\nu_{art}$ with respect to eddy viscosity $\nu_e$ in Figure \ref{fig:QRmesh3}. 

On the coarse meshes M0, M1, and M2, turbulent kinetic energy $k$ is consistently overpredicted, see for instance the profile obtained on M0 (shown in the first row of Figure \ref{fig:QRmesh3}). Increasing the model coefficient $C$ from 0.012 to 0.125 continuously increases both eddy viscosity $\nu_e$ and ensemble artificial viscosity $\nu_{art}$, resulting in larger artificial dissipation that dampens $k$. Consequently, $k$ prediction improves monotonically with the increasing model coefficient. The increased misalignment of the peak position of $k$ with the increasing C is due to laminarization caused by turbulence suppression by the total viscosity. 

On the medium meshes M3, M4, and the fine meshes M5, M6, turbulent kinetic energy $k$ is under-predicted. 
On M3, increasing the model coefficient increases eddy viscosity and artificial viscosity, with artificial viscosity being globally positive. Therefore, on M3, eddy and artificial viscosity together dampen $k$, leading to progressive under-prediction of $k$ with larger model coefficients.  
On M4, the negative artificial viscosity overwhelms the positive part in simulations with $C=0.012$ and $0.048$, while the positive part overwhelms the negative for C=0.101 and 0.125, leading to non-monotonic changes with the change of model coefficient in $k$. However, the overall artificial dissipation on M4 are smaller than the corresponding M3, therefore, high peak in $k$ in the near-wall region.

On the finest mesh M6, $\langle\chi_{nm}\rangle<1$ indicates that the artificial viscosity is globally negative for all coefficients, thereby contributing to the production of $k$. Increasing the model coefficient results in greater dissipation by eddy viscosity and less production by artificial viscosity. However, both $\nu_e$ and $\nu_{art}$ are extremely small (below $3\% \nu$) and counteract each other, resulting in no significant changes in $k$.  

As we can conclude from Figure \ref{fig:QRmesh3}, artificial viscosity can either produce (negative $\nu_{art}$) or dissipate (positive $\nu_{art}$) TKE. Negative artificial viscosity may mimic the backward energy cascade without introducing numerical instability, which is physically realistic and required in many applications. 
artificial viscosity locally changes its role on different meshes. On coarse meshes, artificial viscosity significantly dissipates $k$, and varying the SGS model does not change the sign of net artificial viscosity. 
On medium meshes, artificial error reduces, and varying the SGS model is more likely to change the sign of artificial viscosity, leading to comparative effects of artificial dissipation and production. 
On fine meshes, artificial viscosity becomes extremely small and generally negative, and varying the SGS model still impacts artificial viscosity. Eddy viscosity and artificial viscosity interact and tend to balance each other. 
Accuracy degradation with mesh refinement often occurs when transitioning from medium to fine mesh levels, at a threshold where the global integration of artificial viscosity changes sign. 

\section{Conclusion} 
We have considered the error in LES, focusing on artificial dissipation, based on conservation of turbulent kinetic energy. This method was tested in LES of turbulent channel flow at $\mathrm{Re_\tau}=180$ and 550, using minimum-dissipation model. The comprehensive comparison between our results and DNS results reveals several key insights: 
 
The analysis of time integration methods shows that, under a maximum CFL number of 0.4, first-order temporal schemes introduce significant error, while higher-order schemes demonstrate insensitivity in predictions of flow quantities and viscosities. 
The Forward-Euler method globally under-dissipates energy, resulting in a net production of turbulent kinetic energy and requiring significant eddy dissipation to stabilize the simulation. In contrast, the Backward-Euler scheme over-dissipates energy, leading to excessive suppression of the subgrid-scale (SGS) model. 

All time integration schemes can provide accurate flow predictions with an appropriately chosen time step. For very small time steps, the Forward-Euler scheme achieves better accuracy and computational efficiency than the Crank-Nicolson scheme. However, Crank-Nicolson offers a superior balance between accuracy and efficiency at larger time steps, making it more appropriate for large-scale industrial simulations.
 
Van Kan's pressure correction method, combined with the Backward-Euler scheme, improves the accuracy of all flow quantities, while Chorin's method struggles to produce accurate predictions. For higher-order time integration schemes, both Chorin's and Van Kan's methods yield consistent results across different temporal schemes. Chorin's method demonstrates a stronger pressure transport mechanism, redistributing more turbulent kinetic energy from the streamwise ($u'$) to the wall-normal ($v'$) and spanwise ($w'$) components, leading to improved predictions of $v'$ and $w'$. 

Analysis of spatial discretization schemes shows that the most accurate results are obtained with symmetry-preserving scheme.  Non-conservative schemes result in lower eddy viscosity ($\nu_e < 12\%\nu$), while the artificial viscosity significantly exceeds the eddy viscosity ($5 < \chi_{nm} < 20$). As a result, the LES model is heavily suppressed when spatial discretization schemes introduce excessive artificial errors.  

The insensitivity of $\chi_{nm}$ to changes in the SGS model coefficient indicates that both eddy and artificial viscosities vary similarly, with the eddy viscosity having a dominant influence on the artificial viscosity. artificial viscosity varies its role with mesh refinement: on coarse meshes, it significantly dissipates TKE, while on medium meshes, it can act to produce TKE when the SGS model is adjusted. On fine meshes, the contributions from artificial error and the SGS model become minimal and tend to balance each other. A degradation in the accuracy of turbulence metrics, such as $u'v'$ and $v'$, is observed when transitioning from medium to fine meshes, coinciding with a sign change in the global integration of artificial viscosity.

The most accurate predictions of Reynolds shear stress $u'v'$ across various meshes are achieved when the non-molecular dissipation is close to zero. For accurate predictions of the mean flow kinetic energy ($\bar{E}$), larger non-molecular dissipation is required on finer meshes. For turbulent kinetic energy ($k$), the required amount of non-molecular dissipation decreases as the grid resolution increases.

In summary, the proposed quantification method provides a framework for studying and controlling artificial errors in LES. Insight into the interaction between artificial and eddy viscosities is critical for improving the accuracy of LES predictions.

%%%%%%%%%%%%%%%%%%%%%%%%%%%%%%%%%
%   For journal paper
\section*{Acknowledgements}
This work was supported by the Chinese Scholarship Council and the University of Groningen under Grant (CSC NO. 202006870021). The computing time was granted by the Dutch Research Council (NWO) under project 2022.009.

%Verstappen and Sun express their gratitude to J. Hopman for making the RKSymFOAM and TKE budgets codes publicly available. They also appreciate the valuable discussions with F.X. Trias and Ed Komen.
 
%
%\section*{Disclosure Statement}
%The authors report there are no competing interests to declare.   
 %%%%%%%%%%%%%%%%%%%%%%%%%%%%%%%%%%%%%%%%
\section*{Data availability}
The used code is published on GitHub \cite{sunQR}. The numerical data for the channel flow can be made available upon request.
\bibliographystyle{elsarticle-num.bst} 
\bibliography{abbr_sources}
\end{document}